\newif\iffigs\figstrue

\documentstyle[12pt]{article}
\iffigs
\else
\fi

\batchmode
\newfont{\footscrfont}{rsfs10}
  \newfont{\footbbbfont}{msbm10}
  \newfont{\manfont}{manfnt}
\errorstopmode

\newif\ifscrf\scrftrue
\ifx\footscrfont\nullfont
  \scrffalse
\fi

\newif\ifamsf\amsftrue
\ifx\footbbbfont\nullfont
  \amsffalse
\fi

\def\ppnumber{\vbox{\baselineskip14pt\hbox{hep-th/9712023} 
\hbox{CU-TP-870}
\hbox{HUTP/97-A050}
\hbox{CLNS 97/1524 }}}
\def\ppdate{December 1997}
\def\pplogo{\vbox{\kern-\headheight\kern -15pt
\halign{##&##\hfil\cr&{
\ppnumber}\cr\rule{0pt}{2.5ex}&\ppdate\cr}
}}

\makeatletter
\date{}
\def\dedicatory#1{\def\@date{\normalsize\it#1}}
\def\subjclass#1{\def\@thefnmark{}\@footnotetext{1991
    {\it Mathematics Subject Classification.} #1}}
\def\keywords#1{\def\@thefnmark{}\@footnotetext{
    {\it Key words and phrases.} #1}}

\def\ps@firstpage{\ps@empty \def\@oddhead{\hss\pplogo}%
  \let\@evenhead\@oddhead 
}
\def\maketitle{\par
 \begingroup
 \def\thefootnote{\fnsymbol{footnote}}
 \def\@makefnmark{\hbox
 to 0pt{$^{\@thefnmark}$\hss}}
 \if@twocolumn
 \twocolumn[\@maketitle]
 \else \newpage
 \global\@topnum\z@ \@maketitle \fi\thispagestyle{firstpage}\@thanks
 \endgroup
 \setcounter{footnote}{0}
 \let\maketitle\relax
 \let\@maketitle\relax
\gdef\@thanks{}\gdef\@author{}\gdef\@title{}\let\thanks\relax}

\def\abstract{\if@twocolumn
\section*{Abstract}
\else \small
\begin{center}
{\bf ABSTRACT}
\end{center}
\quotation
\fi}

\def\thebibliography#1{\section*{References\@mkboth
 {REFERENCES}{REFERENCES}}\small\list
 {[\arabic{enumi}]}{\settowidth\labelwidth{[#1]}\leftmargin\labelwidth
 \advance\leftmargin\labelsep
 \usecounter{enumi}}
 \def\newblock{\hskip .11em plus .33em minus .07em}
 \sloppy\clubpenalty4000\widowpenalty4000
 \sfcode`\.=1000\relax}

\newif\iffn\fnfalse

\@ifundefined{reset@font}{\let\reset@font\empty}{} 
\long\def\@footnotetext#1{\insert\footins{\reset@font\footnotesize
    \interlinepenalty\interfootnotelinepenalty
    \splittopskip\footnotesep
    \splitmaxdepth \dp\strutbox \floatingpenalty \@MM
    \hsize\columnwidth \@parboxrestore
   \edef\@currentlabel{\csname  
p@footnote\endcsname\@thefnmark}\@makefntext
    {\rule{\z@}{\footnotesep}\ignorespaces
      \fntrue#1\fnfalse\strut}}}

\makeatother

\ifamsf
  \newfont{\bigbbbfont}{msbm10 scaled\magstep2}
  \newfont{\bbbfont}{msbm10 scaled\magstep1}  
  \newfont{\smallbbbfont}{msbm8}
  \newfont{\tinybbbfont}{msbm6}
  \newfont{\smallfootbbbfont}{msbm7}
  \newfont{\tinyfootbbbfont}{msbm5}
  \newfont{\biggthfont}{eufm10 scaled\magstep2}
  \newfont{\gthfont}{eufm10 scaled\magstep1}  
  \newfont{\smallgthfont}{eufm8}
  \newfont{\tinygthfont}{eufm6}
  \newfont{\footgthfont}{eufm10}
  \newfont{\smallfootgthfont}{eufm7}
  \newfont{\tinyfootgthfont}{eufm5}
\fi

\ifscrf
  \newfont{\scrfont}{rsfs10 scaled\magstep1}  
  \newfont{\smallscrfont}{rsfs7}
  \newfont{\tinyscrfont}{rsfs7}
  \newfont{\smallfootscrfont}{rsfs7}
  \newfont{\tinyfootscrfont}{rsfs7}
\fi

\ifamsf
  \newcommand{\Bbb}[1]{\iffn
      \mathchoice{\mbox{\footbbbfont #1}}{\mbox{\footbbbfont #1}}
      {\mbox{\smallfootbbbfont #1}}{\mbox{\tinyfootbbbfont #1}}\else
      \mathchoice{\mbox{\bbbfont #1}}{\mbox{\bbbfont #1}}
      {\mbox{\smallbbbfont #1}}{\mbox{\tinybbbfont #1}}\fi}
  
\else
  \def\bigbbbfont{\bf}
  \def\Bbb{\bf}
  
\fi

\ifscrf
  \newcommand{\Scr}[1]{\iffn
    \mathchoice{\mbox{\footscrfont #1}}{\mbox{\footscrfont #1}}
    {\mbox{\smallfootscrfont #1}}{\mbox{\tinyfootscrfont #1}}\else
    \mathchoice{\mbox{\scrfont #1}}{\mbox{\scrfont #1}}
    {\mbox{\smallscrfont #1}}{\mbox{\tinyscrfont #1}}\fi}
\else
  \def\Scr{\cal}
\fi

\def\tablerule{\noalign{\hrule}}
\def\operatorname#1{\mathop{\rm #1}\nolimits}
\def\C{{\Bbb C}}

\def\P{{\Bbb P}}
\def\W{{\Bbb W}}

\def\Z{{\Bbb Z}}

\def\Pic{\operatorname{Pic}}

\def\rank{\operatorname{rank}}
\def\bearray{\begin{eqnarray}}
\def\eearray{\end{eqnarray}}
\def\bearraynn{\begin{eqnarray*}}
\def\eearraynn{\end{eqnarray*}}
\def\bfig{\begin{figure}}
\def\efig{\end{figure}}

\def\opeq#1{\advance\lineskip#1 \advance\baselineskip#1
	\advance\lineskiplimit#1}

\def\cM{{\Scr M}}

\def\cD{{\Scr D}}

\def\cMc{{\hfuzz=100cm\hbox to 0pt{$\;\overline{\phantom{X}}$}\cM}}
\def\barcD{{\hfuzz=100cm\hbox to 0pt{$\;\overline{\phantom{X}}$}\cD}}

\ifamsf

\else

\fi

\newtheorem{Proposition}{Proposition}[section]

\newcommand{\be}{\begin{equation}}
\newcommand{\ee}{\end{equation}}

\begin{document}

\title{ $F$-theory and Linear Sigma Models}

\author{M.~Bershadsky$^{1,a}$, T.~M.~Chiang$^{2,b}$, B.~R.~Greene$^{3,c}$,\\
A.~Johansen$^{1,d}$ and C.~I.~Lazaroiu$^{3,e}$}

\maketitle

\vbox{
\centerline{$^1$Lyman Laboratory of Physics}
\centerline{Harvard University}
\centerline{Cambridge, MA 02138}
\medskip
\centerline{$^2$Newman Laboratory of Nuclear Studies}
\centerline{Cornell University}
\centerline{Ithaca, N.Y. 14850}
\medskip
\centerline{$^3$Departments of Physics and Mathematics}
\centerline{Columbia University}
\centerline{N.Y., N.Y. 10027}
\medskip
\bigskip
}

\abstract{
We present an explicit method for translating between  the linear sigma
model and the spectral cover description of $SU(r)$ stable bundles
over an elliptically fibered Calabi-Yau manifold.
We use this to investigate
the $4$-dimensional duality between $(0,2)$ heterotic and $F$-theory
compactifications. We indirectly find that much interesting heterotic
information must be contained in the `spectral bundle' and
in its dual description as a gauge theory on multiple $F$-theory
$7$-branes.
 A by-product of these efforts is a method for analyzing
semistability and the splitting type of vector bundles over an elliptic curve 
given as the sheaf cohomology of a monad.}

\vskip 0.4in

$^a$ bershad@string.harvard.edu

$^b$ chi@lnsth1.lns.cornell.edu

$^c$ greene@phys.columbia.edu

$^a$ johansen@string.harvard.edu

$^e$ lazaroiu@phys.columbia.edu

\pagebreak

\section{Introduction}
Heterotic string theory on Calabi-Yau $3$-folds has long
constituted  a phenomenologically interesting, yet technically
difficult, class  of string compactifications.
Such compactifications
naturally lead to chiral matter in the low energy effective action  
as well
as $N=1$  supersymmetry in spacetime, which underscores their
phenomenological potential. The general such compactification has   
$(0,2)$
world sheet supersymmetry. The lack of left moving
supersymmetry severely hampers a direct perturbative study of such models
and has been the subject of much subtle and inconclusive analysis  
in the past.

The beautiful construction \cite{Witten_phases} of $(0,2)$  
heterotic models
as
infrared renormalization fixed points of certain $(0,2)$ linear  
sigma models
was initially proposed as a controlled method for building $(0,2)$  
nonlinear
sigma models with a fair chance of yielding perturbatively well defined
theories.
Due to its simple and direct connection with the formalism of toric
geometry,
it has enabled a rather explicit study of this accessible class of  
heterotic
$(0,2)$ compactifications.

	The recent nonperturbative string revolution offers the potential
of much new insight into both the formal and phenomenological
aspects of these models. In particular,
$F$-theory gives an alternate description of (some of) these models in
essentially geometrical terms, thereby circumventing many subtle
features of the heterotic formulation. Specifically,
the well known topological and geometrical properties of
Calabi-Yau $4$-folds (which are elliptically and $K3$ fibered)
are thought to encompass the data of an elliptically fibered
Calabi-Yau $3$-fold together with a stable, holomorphic
$E_8 \times E_8$ vector bundle.
The heterotic --- \mbox{\em $F$-{\em theory}} duality
implies that a nonperturbative formulation of $(0,2)$ heterotic
compactifications should cure the perturbative problems which many such
theories apparently exhibit. In particular,
$F$-theory gives an alternate description of these models in essentially
geometrical terms. If the duality of $F$-theory and heterotic
compactifications is taken at face value, one reaches the conclusion that
essentially any \footnote{An exception to this rule is given by heterotic
compactifications whose $F$-theory dual may generate a superpotential
\cite{Witten_superpotential}}
stable $E_8 \times E_8$ vector bundle over an elliptically
fibered Calabi-Yau manifold should lead to a well-defined heterotic model
once one properly includes the full quantum effects.
In this context, $F$-theory gives us an unexpected way of exploring the
full moduli space of $(0,2)$ heterotic compactifications, including its
nonperturbative components.
In fact, in the spirit
of most duality conjectures, this duality is thought
to extend to the full nonperturbative structures on
each side, including various extended brane configurations
that can arise. We see that the potential for progress in
understanding chiral $N=1$ compactifications is considerable, if the map
between their $F$-theory and heterotic descriptions can be made
precise.

A big step in this direction was taken in the beautiful paper of
Friedman, Morgan and Witten \cite{FMW}, where an explicit  
conjecture for the
map from a certain part of the heterotic bundle data to the dual  
$F$-theory
data was proposed. At the same time, three
powerful methods for constructing stable heterotic bundles were explained
and developed. In this paper we will concentrate on the best understood
of these -- the spectral cover description. As explained in \cite{FMW},
\cite{BJPS}, this equivalent description of the heterotic bundle proves
to be extremely useful in analyzing $(0,2)$ compactifications.

From an abstract perspective, giving an $SU(r)$ heterotic
bundle $V$ of rank $r$
in a certain component of the moduli space of stable bundles over an
{\em elliptically fibered} threefold is equivalent to giving a pair
$(\Sigma,L)$ where $\Sigma$ is an $r$-fold cover of the base of the
fibration and $L$ is a line bundle over $\Sigma$.
In heterotic compactifications
on a Calabi-Yau $3$-fold, $\Sigma$ will be a complex surface,
called the spectral cover associated to $V$ due to its origins in  
the theory
of algebraically integrable systems. Beyond giving an equivalent  
description
of a large class of stable bundles (thus giving an effective way to  
implement
the stability condition), this device has proven very useful in  
constructions
of moduli spaces of stable vector bundles over elliptic fibrations
\cite{FMW_math}. Moreover,
as explained in \cite{FMW,BJPS}, this description of
the heterotic bundle
allows one to make rather precise duality statements between certain
heterotic
bundle moduli and corresponding moduli in $F$-theory.

The purpose of this paper is to connect the explicit description
of the only well-understood perturbative heterotic $(0,2)$  
compactifications
--- those realized via $(0,2)$ linear sigma models ---
to the comparatively abstract spectral cover description of  
\cite{FMW,BJPS}.
When combined with the work of \cite{FMW}, this will give us  
a 
framework for understanding certain aspects of the duality in a 
computationally accessible class of models. Going through this  
analysis we
will discover some rather surprising facts, indicating that there
are still some gaps in our understanding of the proposed duality.

In section 2 we briefly review the general form of
$F$-theory---heterotic duality,
as well as the spectral cover construction.
In section 3 we review the linear
sigma model construction of heterotic compactifications.
In section 4 we present an explicit method for recovering the spectral
cover of an $SU(r)$ bundle built via a $(0,2)$ linear sigma model.
In order to avoid unnecessary mathematical complications, only the  
`generic'\footnote{This is the case when the restriction of the 
heterotic bundle to the generic elliptic fibre of the heterotic 3-fold splits 
as a direct sum of line bundles. This condition {\em does} fail to hold in a 
series of examples we constructed.}
case is studied. A complete mathematical discussion of the problem (including 
the `nongeneric' case) 
will be given elsewhere \cite{splitting_type}.
In section 5 we investigate a number of examples. We discover that the
spectral cover of the simplest {\em purely perturbative} models  
that we are
able to construct has a degenerate form.
In section 6 we briefly discuss the
region of the moduli space to which  we think our models belong, leaving
a more
detailed investigation of these issues to be reported elsewhere.
Finally, in section 7 we offer our conclusions and speculations.

While no detailed
understanding of the more subtle aspects of the theory of spectral  
covers are
needed for reading this paper, we do assume a general
familiarity with
both the spectral cover construction
 and with the toric description of $(0,2)$ linear sigma models.
An appendix summarizing certain basic results on stable vector 
bundles over an
elliptic curve has been included for the convenience of the  
non-expert reader.

\section{The geometric framework of $F$-theory- Heterotic duality }

In the original $F$-theory paper \cite{V} a duality between
the heterotic string on $T^2$ and the type IIB string on
$\P^1$ was convincingly advocated. By giving the axion-dilaton of
type IIB a geometric interpretation as a toroidal modulus
(which is at least a consistent
interpretation due to $SL(2,\Z)$ invariance of type IIB string theory),
this duality was rephrased in terms of $F$-theory on an elliptically
fibered $K3$ being dual to the heterotic string on $T^2$.

Once one believes this duality, it is a small step to
generalize it to theories with an ever smaller number of noncompact
dimensions. One does this by making use of the adiabatic
construction of \cite{VW}. Namely, if $B_H$ is some  large volume
manifold then we can use the eight-dimensional $F$-theory/heterotic
duality fiberwise over $B_H$. The idea here is that, as long as $B_H$ is
large,
the compactification over $T^2 \rightarrow B_H$ can be approximated with
a family of $T^2$ compacifications, parametrized by $B_H$. Note, however,
that continuing from this rather trivial situation to the case of a  
base of
finite volume is not straightforward and necessitates a proper  
reformulation
of the conjectures.
Thus, the heterotic string compactified
on $T^2$ fibered over $B_H$ should be isomorphic to
$F$-theory compactified a $K3$ fibration over $B_H$, each of whose  
$K3$ fibers
are themselves elliptically fibered over a $\P^1$ base. The total  
space of this
$K3$-fibration is therefore also elliptically fibered over a base $B_F$,
where $B_F$ is itself a $\P^1$ bundle over $B_H$, i.e. a ruled variety.
The duality between $F$-theory and heterotic models is most
easily expressed when each side is presented in Weirstrass form,
as we shall discuss below.

Since the heterotic construction also involves
specifying a gauge bundle, the data on the heterotic
side generally takes the following form.
We consider an eliptically fibered
$n$-fold (where $n=2,3$) $Z$ over an $n-1$ -dimensional
base $B_H$, whose generic elliptic fiber is denoted  by $E_H$
together with the natural projection map $\pi_H:Z\rightarrow B_H$.
We then specify
$2$ (semistable) bundles
$V$ and $V'$ that control the representations of the left-moving
fermions, thereby specifying the spacetime gauge structure.

As shown in \cite{VM} ,
the correct dual of this is an $(n+1)$-fold $X$, elliptically fibered over a  
base $B_F$
by a map $\pi_F:X\rightarrow B_F$,
where $B_F$ is a ruled $n$ fold over $B_H$
(simply speaking, a $\P^1$-bundle over $B_H$).
The generic elliptic fibre of $X$ will be denoted by $E_F$.
$B_F$ can be obtained as the projectivisation
\footnote{If $W$ is a vector bundle
over a base $B$, then the projectivisation of $W$ is most easily  
defined in
terms of complex geometry, as the variety
$\P(W)=W/\C^*$ where $\C^*$ is the natural fiberwise action
$\rho :\C^* \rightarrow End(W)$ given by:
$\rho(\lambda)(u):=\lambda u, \forall u \in E_b, \forall b \in B$.
Intutively, we associate to each fiber $W_b$ of $W$ the projective space
$\P(W_b)=W_b/\C^*$ and we glue these together in a
$\P^{\rank W - 1}$-fibration
over $B$ in a way specified by the fiberwise character of the  
action $\rho$.
If $W=\oplus_{a=1..m}{L_a}$ is a sum of line bundles, then a natural
generalization of this construction is to define
$\P_{w_1...w_m}(W):=W/\rho_{w_1...w_m}$, 
where $\rho_{w_1...w_m}$ is the natural
fiberwise $\C^*$ action on $W$ given by:
$\rho_{w_1..w_m}(u_1\oplus...\oplus u_m):=\lambda^{w_1}u_1\oplus...\oplus
\lambda^{w_m}u_m$, the weights $w_1..w_m$ being
positive integers. Then $\P_{w_1...w_m}(W)$ is a weighted  
projective space
fibration over $B_H$, which will in general have singularities and  
is thus
only a variety and not a manifold. The case $w_1=...=w_m=1$ leads to the
usual projectivisation above. $\P_{w_1...w_m}(W)$ is equiped with the
{\em twisting sheaf }  $O_{\P_{w_1...w_m}(W)}(1)$, which is a line  
bundle in
the
case $w_1=...=w_m$. $O_{\P_{w_1...w_m}(W)}(1)$ has the property that its
restriction to each fibre $O_{\P_{w_1...w_m}}(W)_b \approx
\W\P_{w_1...w_m}^{m-1}$
is isomorphic with the twisting sheaf  $O_{\W\P_{w_1...w_m}^{m-1}}(1)$
of $\W\P_{w_1...w_m}^{m-1}$.
Moreover, there are sections
$x_i \in H^0(L_i \otimes O_{\P_{w_1...w_m}(W)}(w_i))$ (called   
homogeneous
coordinates of $\P_{w_1...w_m}(W)$), which restrict on each
fibre to usual homogeneous coordinates of $\W\P_{w_1..w_m}^{m-1}$.
The reader can consult \cite{Hartshorne,Griffith_Harris,Grothendieck}
for more details on this basic construction in algebraic geometry.}
$B_F:= \P({\cal M}\oplus {\cal O}_B)$,
where the line bundle ${\cal M} \rightarrow B_H$
is (partially) determined
by the characteristic classes of $V, V'$ as explained in \cite{FMW}.
This specifies the $\P^1$-fibration structure of $B_F$, allowing us to
reconstruct $B_F$ from the data of $B_H$ and $V,V'$.
The Kahler moduli of $B_H$, together with the expectation value of the
heterotic dilaton, map to the Kahler moduli of $B_F$
(which are the only Kahler moduli visible in $F$-theory on $X$).

The bundle data  $V, V'$ are equivalent to pairs $(\Sigma, L)$,  
respectively
$(\Sigma',L')$, where $\Sigma, \Sigma'$ are $n-1$-folds which are
generically $r$-branched covers of $B_H$ and $L, L'$ are ---  
generically ---
line
bundles over $\Sigma,\Sigma'$.  $\Sigma$ and $L$ are called the
{\em spectral cover} and the {\em spectral bundle} of $V$, as will
be discussed in more detail below.


The important point is
that this equivalent description of $V, V'$ allows us to organize their
moduli in a way which facilitates describing
the  map to the moduli of the F-theory dual on $X$.
Namely, the complex structure of $Z$ together with the complex structure
moduli of
$\Sigma, \Sigma'$ and the Kahler modulus of the generic elliptic fibre of
$Z$ map to the complex structure of $X$, while the moduli of $L$ map to
certain RR moduli of $F$-theory on $X$.

To make these statements more precise, one follows \cite{V,VM,BIKMSV}
in writing both $X$ and $Z$ as Weierstrass models
\footnote{An elliptic fibration $Y \longrightarrow B$ with a section
$\sigma$
will typically have singular fibers
even if $ Y $ itself is smooth. Some of these singular fibers will  
simply be
elliptic curves with a node (so called type $I_1$) or
with a cusp (type $II$),
but in general there will also be reducible fibres.
The Weierstrass model $Y_W$ of $Y$ is obtained from $Y$ by blowing  
down all
components of the reducible fibers which do not meet $\sigma$. This will
typically introduce extra singularities in $Y$.
$Y$ can be recovered from $Y_W$ by blowing up these singularities  
(a process
called a { \em small resolution}). In our case, the Kahler parameters
associated
to small resolutions of $X_W$ are immaterial since they are not physical
parameters in $F$-theory, but the Kahler parameters of small resolutions
of $Z_W$ are physical. Therefore, using a Weierstrass model for $Z$ is,
strictly speaking, only
possible over some locus in the moduli space of the heterotic  
theory on $Z$
where these moduli are set to zero. On the $F$-theory side this
corresponds to a particular locus in the complex
structure moduli space of $X$. However, throughout this paper, we  
will only
be
interested in the case when $Z$ is smooth, so its only degenerate fibers
are of types $I_1$ and/or $II$, which are irreducible and necessitate no
blow-down.
In this case, $Z$ is already a Weierstrass model. In other words,  
we do not
allow
ourselves to explore the full moduli space of our theories but only a
subspace of it.}.
For this, let $K_{B_H}$ and $K_{B_F}$ be the canonical line
bundles of
$B_H$ and $B_F$. (Due to the ruled character of $B_F$, it is easy  
to see that
we
have $K_{B_F}^{-1}=K_{B_H}^{-1}\otimes O_F(2)\otimes {\cal M}$,
where $O_F(1)$ is  the
twisting sheaf of $ B_F=\P({\cal M}\oplus {\cal O}_{B_H})$ ).
Then one constructs the projectivisations:
$\P_H:=\P_{2,3,1}(K_{B_H}^{-2}\oplus K_{B_H}^{-3} \oplus {\cal O}_{B_H})$
(over $B_H$)
and
$\P_F:=\P_{2,3,1}(K_{B_F}^{-2} \oplus K_{B_F}^{-3} \oplus {\cal  
O}_{B_F})$
(over $B_F$)
where the weights of the natural $\C^*$ actions are indicated by  
subscripts.
If $x \in H^0(K_{B_H}^{-2} \otimes O_{\P_H}(2)),y \in
H^0(K_{B_H}^{-3}\otimes O_{\P_H}(3)),z \in H^0(O_{\P_H}(1))$,
respectively
$X \in H^0(O(K_{B_F}^{-2}\otimes O_{\P_F}(2)),Y
\in H^0(K_{B_F}^{-3}\otimes O_{\P_F}(3)),Z \in H^0(O_{\P_F}(1))$
are the homogeneous coordinates of the projectivisations $\P_F,\P_H$,
then we can write $Z$,$X$ as
the  zero divisors associated to the sections
\be
\hskip -.2in {\rm heterotic:} ~~~~
y^2 - x^3-fxz^4 -gz^6 \in H^0(K_{B_H}^{-6} \otimes O_{P_H}(6))
\ee
respectively
\be
{\rm F-theory:} ~~~~
Y^2 -X^3 -FXZ^4 -GZ^6 \in H^0(K_{B_F}^{-6} \otimes O_{\P_F}(6)),
\ee
where $f \in H^0(K_{B_H}^{-4}),g \in H^0(K_{B_H}^{-6})$
and $F \in H^0(K_{B_F}^{-4}),G \in H^0(K_{B_F}^{-6})$
are sections specifying the complex structure of $Z,X$.
Since $B_F$ is a $\P^1$  bundle over $B_H$, one can make an
expansion in the coordinate of the $\P^1$ fibre.
Geometrically, this is described as follows. Let
$u \in H^0({\cal M} \otimes O_{F}(1)) ,
v \in H^0({\cal O}_{B_H} \otimes O_{F}(1))$
be the
homogeneous coordinates of the projectivisation
$B_F=\P({\cal M} \oplus {\cal O}_{B_H})$. Then we can expand:
\be
\label{F_expansion}
F=\sum_{i=0..I}{u^i \otimes v^{8-i}\otimes F_i}
\ee
\be
\label{G_expansion}
G=\sum_{j=0..J}{u^j \otimes v^{12-j}\otimes G_j}
\ee
where $F_i \in H^0(K_{B_H}^{-4} \otimes {\cal M}^{4-i})$,
$G_j \in H^0(K_{B_H}^{-6} \otimes {\cal M}^{6-j})$ and
$I,J$ are the largest values of $i,j$ for which
$H^0(K_{B_H}^{-4} \otimes {\cal M}^{4-i})$, respectively
$H^0(K_{B_H}^{-4} \otimes {\cal M}^{4-i})$, are nonzero.

The work of \cite{VM,BIKMSV,FMW,BJPS}
then shows that the complex structure
moduli of $Z$ (which are controlled by $f$ and $g$) map to the moduli
controlled by the `middle sections' \footnote{Here we assume that  
there is
enough amplenesss of $K_{B_H}^{-1}$,which is usually the case.
In $4$ dimensional
compactifications,one usually takes $B_H$ to be a Fano $2$-fold, so
$K_{B_H}^{-1}$ is ample.}
$F_4$ and $G_6 $, while the moduli of $\Sigma$ and $\Sigma'$ map to
the moduli of the `lower sections' $F_i, G_j$  $( i <4 ,j <6 )$,  
respectively,
of the
`upper sections' $F_i,G_j $  $( i >4 ,j >6 )$
(one can formally extend the sums
above up to $i=8$, respectively $j=12$, by defining $F_i,G_j$ to be
identically zero for $i>I$, respectively for $j>J$; this explains the
terminology `middle sections').

For the sake of the non-expert reader let us specialize this to
the simplest case of six-dimensional compactifications
\footnote{For a 4-dimensional example see section 6.1.}
with $B_H = \P^1$,
for which the choice ${\cal M}=O_{\P^1}(m)$ leads to $B_F$ being  
the Hirzebruch surface $F_{m}$.
As a toric variety, $F_m=(\C^4-{\cal F})/(\C^*)^2$
where ${\cal F}$ is the exceptional set  and the holomorphic  
quotient action
is given by
$(s,t,u,v)\stackrel{(\lambda,\mu) \in (\C^*)^2}{\longrightarrow}
(\lambda s,\lambda t,\lambda^m \mu u,\mu v)$.
The homogeneous coordinates $s,t,u,v$ are thus sections of the line  
bundles
$O(1,0),O(1,0),O(m,1),O(0,1)$, respectively.
The coordinate pairs $(u,v)$,
$(s,t)$ can be viewed as the homogeneous coordinates of two $\P^1$'s.
Since $(s,t)$ are left invariant by the action of $\mu$, $F_m$ as a ruled
surface is a $\P^1$-fibration over $\P^1$.
The coordinates
$(s,t)$ are associated with the $\P^1$ base (denoted by $\P^1_b$) while
the coordinates
$(u,v)$ are associated with the $\P^1$ fibre (denoted by  
$\P^1_f$). In this case
$\P^1_b$ is nothing other than $B_H$.
Clearly, this $\P^1$-fibration is just the projectivisation
$F_{m}=\P(O_{\P^1_b}(m)\oplus {\cal O}_{\P^1_b})$, so that
${\cal M}=O_{\P^1_b}(m)$ in this case.
The twisting sheaf $O_{F}(1)$
of this projectivisation can be naturally identified with $O_{F_m}(0,1)$.
Thus $O_{\P^1_b}(m) \otimes O_F(1)\approx O_{F_m}(m,1)$,
 ${\cal O}_{\P^1_b} \otimes O_F(1)\approx O_{F_m}(0,1)$ and  
$(u,v)$ can be
naturally identified with homogeneous coordinates of the  
projectivisation.
A canonical divisor of $F_m$ is given by $ -(D_u+D_v+D_s+D_t) \sim 
-2D_v - (m+2)D_s=-(stuv)$
(where $D_u$, $D_v$, $D_s$, $D_t$
are the toric divisors of $F_m$ and $\sim$ denotes linear equivalence), 
so the anticanonical line bundle is
$K_{F_m}^{-1}=O_{F_m}(m+2,2)$.
Since $B_F$ is a toric variety,
the ambient space
of the Weierstrass model of $X$ is itself toric and given by:
$\P_F:=\P_{2,3,1}(K_{F_m}^{-2}\oplus K_{F_m}^{-3}\oplus {\cal O}_{F_m})
\approx
(\C^7-{\cal F}')/(\C^*)^3$, with homogeneous coordinates
$(s,t,u,v,X,Y,Z)$ and $(\C^*)^3$ action :
\be
(s,t,u,v,x,y,z)\stackrel{(\lambda,\mu,\nu) \in (\C^*)^3}{\longrightarrow}
(\lambda s, \lambda t, \lambda^m \mu u ,\mu v ,\lambda^{2m+4}\mu^4  
\nu^2 X,
\lambda^{3m+6}\mu^6 \nu^3 Y,\nu Z).
\ee
By the same reasoning as above, the twisting sheaf of this  
`projectivisation'
is $O_\P(0,0,1)$ and its homogeneous coordinates $X,Y,Z$
are sections of $O_{\P_F}(2m+4,4,2),O_{\P_F}(3m+6,6,3)$ and
$O_{\P_F}(0,0,1)$.
Then $F \in H^0(O_{\P_F}(4m+8,8,0))$, $G \in H^0(O_{\P_F}(6m+12,12,0))$  
can be viewed
as polynomials $F=F(s,t,u,v)$, $G=G(s,t,u,v)$
in $(s,t,u,v)$ of multidegrees  $(4m+8,8)$, $(6m+12,12)$ under the  
action of
$(\lambda,\mu)$, while $F_i \in H^0(O_{\P^1_b}(m(4-i)+8))\approx 
H^0(O_{\P_F}(m(4-i)+8,0,0))$, 
$G_j \in H^0(O_{\P^1_b}(m(6-j)+12))\approx H^0(O_{\P_F}(m(12-j)+12,0,0))$ can be identified with polynomials 
in $(s,t)$ of degrees $m(4-i)+8$, respectively $m(6-j)+12$ under the action 
of $\lambda$.
On the other hand, the ambient space of the Weierstrass model of $Z$ is 
the toric variety 
$\P_H=\P_{2,3,1}(K_{\P^1_b}^{-2}\oplus K_{\P^1_b}^{-3}\oplus 
{\cal O}_{\P^1_b})\approx (C^5-{\cal F''})/(C^*)^2$ with homogeneous 
coordinates $(s,t,x,y,z)$ and $(C^*)^2$ action :
\be
(s,t,x,y,z) \stackrel{(\lambda,\rho) \in (C^*)^2}{\longrightarrow}
(\lambda s, \lambda t, \lambda^4\rho^2 x, \lambda^6 \rho^3 y, \rho y)
\ee
and $f \in H^0(O_{\P^1_b}(8))\approx H^0(O_{\P_H}(8,0)), 
g \in H^0(O_{\P^1_b}(12))\approx H^0(O_{\P_H}(12,0))$ can be viewed 
as homogeneous polynomials $f(s,t), g(s,t)$ of degree 8, respectively 12 
under the action of $\lambda$. This recovers the more intuitive
(albeit less geometric) description of \cite{VM,BIKMSV}.

A precise asymptotic form of the map from $\Sigma$, $\Sigma'$ to the
lower/upper
sections (in the case when the low energy effective theory has enhanced
gauge symmetry)
was conjectured in \cite{FMW},
but the precise map between
the complex structure moduli of $Z$ and the middle sections is not yet
understood. Furthermore,
in  general, an $F$-theory compactification on $X$ requires  
specifying not
only
the $n$-fold $X$ but also a certain point $\omega$ in a  
subset
$IJ_c(X)$ of the second intermediate jacobian $IJ_2(X)$ of $X$,  
giving the
corresponding RR moduli of the compactification \cite{FMW}.
In \cite{FMW} it was argued that the moduli of $L, L'$ map to the  
moduli of
this point $\omega \in IJ_c(X)$, but the precise form of this map is
also unknown at present.

Now let us discuss in more detail the correspondence between a stable
bundle $V$ over $Z$ with $c_1(V)=0$ and the pair $(\Sigma,L)$.
Here we assume that $c_2(V)$ and $c_3(V)$ are kept fixed, which  
gives us a
moduli space of stable bundles ${\cal M}(0,c_2,c_3)$.
\footnote{Note that this moduli
space is believed to be stratified in the general case
\cite{FMW,BJPS}}.

Intuitively, the spectral cover $\Sigma$ encodes the data of the  
heterotic
Wilson lines over each toroidal fibre. Mathematically, the  
construction of
$(\Sigma,L)$
 is roughly as follows.
One limits oneself to bundles $V$ with the property that their
restriction to the generic elliptic fibre of $Z$ is a direct sum
$V_E=\oplus_{j=1..r}{L_j}$ of line
bundles over $E$. Note that a `generic' semistable bundle of degree  
zero over
an elliptic curve will have this property and thus one can argue that the
`generic' stable bundle $V$ over $Z$ will belong to this component of the
moduli space. As explained in the appendix, semistability of $V_E$ forces
deg$L_j=0$ for all $j$ and thus $L_j \approx O_E(q_j-p)$ for some points
$q_j \in E$, where $p$ is the distinguished point on $E$ where
$E$ touches the section of the elliptic fibration of $Z$.
If $\pi_H(E)=b \in B_H$, this gives a family of points $q_j(b)$ ($j=1..r$)
with the property that $\pi_H(q_j(b))=b$.
As $b$ varies over $B_H$, this
describes an $r$-fold cover of $B_H$ which we call $\Sigma$, living as a
hypersurface in $Z$.
Due to the compactness of $Z$, this cover will necessarily be  
branched and
the above construction will in general fail over the elliptic  
fibers of $Z$
lying above the branching locus, which generically is a divisor in $B_H$.
In fact,
a correct mathematical construction of $\Sigma$ reveals that
$V_E$ will typically fail to be fully split for $E$ lying above the  
branching
locus, although the corresponding points of
$\Sigma$ can still be obtained by
considering the pieces of the associated graded bundle of $V_E$
(see the appendix).
We see that,
in order to recover $\Sigma$ from $V$,
all we need, in first approximation, is to understand the
splitting behaviour of
$V_E$ for a generic elliptic fibre $E$ of $Z$.
This will be accomplished in
section 4 by studying the properties of the sections of $V_E$.
More precisely, we will show how one can 
recover $\Sigma_{B_H - Q}$ where $Q$ is a divisor 
in $B_H$ above which $V_E$ is not fully split. This leaves us with a 
certain ambiguity (which is potentially related to interesting physics),
a detailed discussion of which will be given elsewhere.

The reason for the 
appearance of $L$ is more subtle and can only be properly understood
in a rigorous construction.
The point is that $\Sigma$ 
does not contain enough information to recover $V$ and one can see that
the missing data are given by a line bundle $L$ over $\Sigma$.  
Intuitively,
$\Sigma$ only fixes the isomorphism class of $V_{E_b}$ for each fiber
$E_b =\pi_H^{-1}(b)$ but not the way in which the {\em concrete}
representative of $V_{E_b}$ `twists' as $b$ varies in $B_H$ in order to
form $V$.
By including $L$, one can show that the correspondence between
$(\Sigma,L)$ and $V$ is one to one.
For more details the reader can consult
\cite{FMW,BJPS,FMW_math,Donagi_Markman}\footnote{The non-expert reader should  
be aware
that our explanation almost completely sacrificed mathematical rigour in
favour of simplicity. For the reader acquainted with \cite{FMW_math}, let us 
mention an important difference between the bundles which will appear in our 
examples in section 5 and the bundles considered there. The most 
general results of \cite{FMW_math} concern bundles of trivial determinant 
whose restriction to the 
generic elliptic fibre of $Z$ is not only semistable but also regular -- 
which means that the line bundles $L_i$ appearing in maximal direct sum 
decomposition 
of $V$ explained in the appendix are 
mutually distinct. In particular, this rules out the tangent bundle of a 
threefold. Most of the 4-dimensional heterotic 
compactifications in section 5 do not satisfy this regularity condition.}.

The cohomology class associated to $\Sigma$ 
can be computed in terms of the characteristic 
classes of $V$ as explained in \cite{FMW}. In general, $\Sigma$ will be given 
by the zero divisor of a section $s =\sum_{2i+3j+k=r}{a_{ijk}x^iy^jz^k}
\in H^0({\cal N} \otimes O_Z(n\sigma))$, 
where ${\cal N}$ is a line bundle over $B_H$ and $\sigma$ is the 
section of the elliptic fibration of $Z$. Here $r:=\rank V$ and  
$a_{ijk}\in H^0({\cal N} \otimes K_{B_H}^{-(n-k)})$, the sum being restricted to 
nonnegative $i,j,k$. By using the equation of $Z$, this can always be written 
in the canonical form: 
\be
s=\sum_{0\le e=even\le r}{a_e x^{e/2}z^{r-e}} + y\sum_{3\le e=odd \le r}
{a_e x^{(e-3)/2}z^{r-e}}
\ee
where $a_e:=a_{e/2,0,r-e}$ for $e$ even and $a_e:=a_{(e-3)/2,1,r-e}$ for $e$ 
odd. We have $a_e \in H^0({\cal N} \otimes K_{B_H}^e)$ for all $e$. 
It was shown in \cite{FMW} that:
\be
c_1({\cal N})=\pi_{H,*}(c_2(V))
\ee
This relation fixes ${\cal N}$ completely if $\Pic^0(B_H)=0$, which is the 
case, in particular, if $B_H$ is a toric variety.
Remember that the dual heterotic $n$-fold is also characterized by the 
line bundle ${\cal M}$, which specifies $B_F$ as a  $\P^1$-fibration 
over $B_H$. 
The conjecture of \cite{FMW} is that ${\cal M}={\cal N}\otimes K_{B_H}^{6}$. 
Moreover (concentrating on one of the $E_8$ factors), if we assume 
that the gauge 
symmetry of the effective theory is not completely broken, then $a_e$ 
should be identified with certain sections related to  
$F_i,~G_j$ which occur naturally once one imposes on $X$ the constraints 
required by Tate's algorithm \cite{BIKMSV}.

In the case of 6-dimensional compactifications or of 4-dimensional 
compactifications with  $B_H$ a  Hirzebruch surface $F_k$, 
but $Z$ otherwise generic, one 
can determine the class of $\sigma$ more explicitly 
as explained in \cite{BJPS}
\footnote{Simplifications occur in this case because of the ruled character 
of $B_H=F_k$, which implies that $Z$ is itself K3-fibered.}. 
Let us recal the relevant results.
Consider a generic stable bundle $V$ with $\det V \approx {\cal O}_Z$ 
over a {\em smooth} $n$-fold $Z$, elliptically fibered over $B_H$ ($n=2,3$).
Let $S:=[\sigma(B_H)]\in H^{2}(Z,\Z)$ be the cohomology 
class of $\sigma(B_H)$ and 
$F\in H^{2n-2}(Z,\Z)$ be the cohomology class of the elliptic fibre of $Z$.
Let $[\Sigma] \in H^{2}(Z,\Z)$ be the cohomology class associated to $\Sigma$.

In the appendix of \cite{BJPS} it is shown that
\footnote{Juxtaposition denotes the cup product in integer cohomology, while 
$c[Z]$ is the evaluation of a cohomology 
class $c \in H^{2n}(Z,\Z)$ on the fundamental 
homology class $[Z] \in H_{2n}(Z,\Z)$ of $Z$.} :

(1) If $Z$ is an elliptically fibered K3 surface with $\Pic(Z)\approx \Z^2$, 
then :

\be
\label{class2}
[\Sigma] = \rank V ~S + c_2(V)[Z]~F 
\ee

(2)If $Z$ is an elliptically fibered Calabi-Yau threefold with
$B_H$ a Hirzebruch surface $F_k$ (with $a,b \in H^{2}(B_H,\Z)$ the 
cohomology classes of the $\P^1$-fibre, respectively of the infinity section 
of $F_k$) 
and if the classes $A:=\pi_H^*(a)$, $B:=\pi_H^*(b)$ and $S:=[\sigma(B_H)]$ 
form a $\Z$-basis of the abelian group $H^2(Z,\Z)$ and 
generate the graded ring 
$H^*_e(Z):=H^0(Z,\Z)\oplus H^2(Z,\Z)\oplus H^4(Z,\Z) \oplus H^6(Z,\Z)$, 
then $A  B$, $A  S$ and $B  S$ form a $\Z$-basis of the group $H^{4}(Z,\Z)$.
Moreover, if $c_2(V)=c_2(V)_{AS}~A  S+c_2(V)_{BS}~B  S+c_2(V)~A  B$ is the 
associated decomposition of $c_2(V)$, then we have :
\be
\label{class3}
[\Sigma] = \rank V ~S + c_2(V)_{AS} ~A+c_2(V)_{BS} ~B 
\ee
Obviously a generic smooth elliptic K3 surface and a generic smooth  
Calabi-Yau threefold elliptically fibered over a Hirzebruch surface satisfy 
the above hypotheses.

Since the  spectral covers $\Sigma$,
$\Sigma'$ play a key role in
the proposed duality map, this naturally
leads us to try to understand how they can be explicitly identified.
Specifically, since $(0,2)$ linear sigma models are our most
insightful tool for directly constructing heterotic models,
we would like to find a procedure for identifying
$\Sigma$,
$\Sigma'$ in terms of linear sigma model data.
To this end, let us briefly recall the linear sigma model construction.

\section{$(0,2)$ Models and the Linear Sigma Model Construction}

The  linear sigma models we consider are abelian supersymmetric gauge
theories in two dimensions with $(0,2)$ supersymmetry.
Their interest stems from the well-known arguments of  
\cite{Witten_phases}
to the effect that, if such a model is carefully defined, then it is
expected to have
an infra-red fixed point whose conformal field theory gives a good  
starting
point for a perturbative heterotic compactification.

Beyond a $U(1)^h$ worldsheet gauge symmetry, such a model has
a number $s$ of local fermionic symmetries. Associated  to these
symmetries one has $h$ abelian gauge multiplets and $s$ pairs
\footnote{The notation we use follows the conventions of \cite{CDG},
to which
we refer the reader for further details. We denote the lowest  
component of a
superfield by the lower case form of the same letter. Thus,  
$\phi_\rho$ is the
lowest component of $\Phi_\rho$ etc.}
$(\Sigma_j,\overline{\Sigma_j})(j=1..s)$ of
(nonchiral) complex fermi superfields (`twisted Fermi superfields').
The worldsheet matter content consists of $d+1$ chiral scalar  
superfields,
denoted $\Phi_\rho (\rho=1..d)$ and $P$ as well as $m+t$ chiral fermi
superfields
denoted $\Lambda_a (a=1..m)$ and $\Gamma_i (i=1..t)$.
One also includes
spectator fields as discussed in \cite{Distler_notes}.
Each of the matter
fields transforms in a  representation of $U(1)^h$
characterized by a multicharge $q \in \Z^h$.
The only fields which transform
nontrivially under the local fermionic symmetries are $\Gamma_i$,  
$\Lambda_a$
and $\Sigma_j$. The transformation laws of $\Gamma_i$ and $\Lambda_a$ are
controlled by holomorphic functions $E_{0i}^{(j)}(\Phi) (i=1..t,j=1..s)$,
respectively $E_a^{(j)}(\Phi) (a=1..m,j=1..s)$. Beyond the usual  
gauge kinetic
and Fayet-Iliopoulos D and theta terms and the kinetic terms of the  
matter
fields, the action contains a term describing a coupling of the fermionic
superfields $\Lambda_a$ to the twisted fermionic fields $\Sigma_j$.  
It also
contains a superpotential which couples
$\Gamma_ i$ to holomorphic combinations $G_a(\Phi_1..\Phi_d)$,
$P$ to $\Lambda_a$ and to holomorphic combinations
$F_a(\Phi_1...\Phi_d)$, as well as spectator field terms.
Our notation for the various $U(1)$ charges is given below :
$$
\hbox{
\vbox{\offinterlineskip \tabskip=0pt
\halign{
#&
\vrule height 10pt depth 5pt
\enskip\hfil$#$\hfil\enskip\vrule &
\enskip\hfil$#$\hfil\enskip\vrule &
\enskip\hfil$#$\hfil\enskip\vrule &
\enskip\hfil$#$\hfil\enskip\vrule &
\enskip\hfil$#$\hfil\enskip\vrule \cr\tablerule&
\phi_\rho&
p&
\lambda_a&
\gamma_i \cr \tablerule &
q_\rho&q_0&{\tilde q}_a&{\tilde q}_{0i}\cr\tablerule
}}
\qquad
\qquad
\vbox{\offinterlineskip \tabskip=0pt
\halign{
#&
\vrule height 10pt depth 5pt
\enskip\hfil$#$\hfil\enskip\vrule &
\enskip\hfil$#$\hfil\enskip\vrule &
\enskip\hfil$#$\hfil\enskip\vrule &
\enskip\hfil$#$\hfil\enskip\vrule &
\enskip\hfil$#$\hfil\enskip\vrule \cr\tablerule&
E_a^{(j)}&
E_{0i}^{(j)}&
F_a&
G_i\cr\tablerule&
{\tilde q}_a&{\tilde q}_{0i}-q_0&-q_0-{\tilde q}_a&-{\tilde
q}_{0i}\cr\tablerule
}}}
$$

A semiclassical study of these models
\cite[and references therein]{CDG,MDG} reveals an intricate phase
structure for their moduli space which can be explicitly analyzed
by methods from toric geometry.
Physically, these phases arise by changing the values of the $h$
Fayet-Iliopoulos parameters $r_i$ on which the models
depend.
Typically one builds such a model so that it admits a Calabi-Yau phase in
which all of the $r$-parameters are positive. By following the  
assignment of
gauge charges for the various fields, one finds out that the moduli space
in such a region is described --- at least in the limit of large
$r$-parameters ---
by a nonlinear $(0,2)$ sigma model
defined over a Calabi-Yau variety $Z$  realized as a
complete intersection in a toric variety.
Such a model is specified,
in the classical limit, by $Z$ and a vector bundle $V$ over $Z$.
To interpret the data $(V,Z)$ as a starting point for a  
perturbative string
compactification, $Z$ must satisfy the Calabi-Yau condition and $V$ must
admit the
structure of a (semi-) stable holomorphic vector bundle over $Z$ with
$c_1(V)=0$, $c_2(V)=c_2(T)$.
This imposes certain
constraints on the gauge charges of the original $(0,2)$ linear  
sigma model,
which turn out to be equivalent to the absence of an anomaly for the
gauge, left $U(1)$ global  and right $R$-symmetries of that theory.
However, the emergence of the stability condition for $V$ is
somewhat mysterious in this context, and very hard to test. At
present, we do not know the full set of conditions
which ensure that a
well-defined $(0,2)$ linear sigma model  gives rise to a (semi-)  
stable bundle
$V$.

In practice, a self-consistent way to test this is provided by the  
analysis
of the linear sigma model in other phases. Concretely, such a model will
typically
also admit a Landau-Ginzburg phase in which it reduces to a
supersymmetric Landau-Ginzburg model. By analyzing the 
corresponding conformal field theory one can extract information  
about the
spectrum \cite{Witten_02_spectrum,Distler_Kachru_LG}.
It is believed that inconsistency of the theory at the
Landau-Ginzburg point signals a failure of the stability condition  
for $V$.
One of the main points that we will make in the following is that,
when we talk about a `purely perturbative' linear sigma model, we will
require it to be well-behaved at the Landau-Ginzburg point. Note  
that, since
we do not have a good direct way to test semistability of $V$ over  
$Z$, we
regard this condition as central for having a well-defined perturbative
model.
Careful implementation of these  consistency conditions will have
a significant effect on the results we find.

We now explain the mathematical description of the model in the deep
Calabi-Yau phase, which --- modulo
the above --- is the only part of the linear $(0,2)$ moduli space
directly entering our subsequent discussion. In the limit of large but
fixed positive
Fayet-Iliopoulos parameters, the moduli space is the simplectic
quotient of the minimal set of the bosonic potential via the action  
of the
gauge group $U(1)^h$. Rewriting this as a holomorphic quotient one  
obtains
a complete intersection $G_1(\Phi)=..=G_t(\Phi)$
in a toric variety $\P=(\C^d-{\cal F})/(\C^*)^h$, where ${\cal F}$ is the
exceptional set.
The bosonic fields $\phi_\rho$ transform as homogeneous coordinates  
of $\P$.
Since in this phase they are restricted to lie in $Z$, we will however
denote the homogeneous coordinates of $\P$ by $x_\rho$, with the
understanding that on $Z$ we have $\phi_\rho=x_\rho$.
Analysis of the massless fermionic modes reveals that
the gauge data $V$ are given by the cohomology
of a monad. Specifically, this means that $V$ is defined by the  
cohomology of
the complex :

\begin{equation}
\label{monad} 0 \longrightarrow \oplus_{j=1}^{s}{O_Z}
\stackrel{f}{\longrightarrow} \oplus_{a=1}^{m}{O_Z(\tilde{q_a})}
\stackrel{g}{\longrightarrow} O_Z(-q_0)\longrightarrow 0
\end{equation}
In this equation, the maps $f$ and $g$ act on the fibers as follows :
\be
f_x(\eta_1 ... \eta_s):=\left( \begin{array}{ccc}  E_1^{(1)}(x) & ... &
E_1^{(s)}(x) \\
					    E_2^{(1)}(x) & ... &  
E_2^{(s)}(x) \\
					    ...     & ... & ...       \\
					    E_m^{(1)} (x) & ... &  
E_m^{(s)}(x)
		    \end{array} \right)
\left( \begin{array}{ccc} 	\eta_1 \\
		   	\eta_2 \\
			... \\
			\eta_s
\end{array} \right)
, \forall x \in Z, \forall \eta_j \in O_{Z,x}
\ee
where $E_a^{(j)} \in H^0(O_Z({\tilde q_a}))$, and
\be
g_x(v_1 ... v_m):=\sum_{a=1..m}{F_a(x)\otimes v_a}
\forall x \in Z, \forall v_a \in O_{Z,x}({\tilde q}_a)
\ee
where $F_a \in H^0(O_Z(-q_0 -{\tilde q_a}))$.
By taking dimensions in (\ref{monad})
one easily sees that we have $m=r+s+1$ where $r:=$rank$V$.
The sections of $V$ are identified with masless linear combinations  
of the
fermionic coordinates $\lambda_a$ .

In order
to avoid complications, we will choose the defining data
so that $V$ is a {\em bundle} over a nonsingular
Calabi-Yau {\em manifold } $Z$.
As argued in \cite{CDG}, it is possible to
consider more general situations, but we will restrict to this case.
In particular, this means that we choose the maps $f,g$ in the monad
to be transverse.

Analysis of the Landau-Ginzburg phase as in \cite{CDG} leads
to the conclusion that the number $s$ of
local fermionic symmetries of the
underlying linear sigma model
(which is equal to the number of columns
$E^{(j)}$ of the matrix above)
must be large enough to avoid having massless charge zero fermions at
the Landau-Ginzburg point.
Mathematically, this is reflected by the fact that,
since $c_1(V)=0$, a necessary condition for stability of $V$ is that
$H^0(V)=0$.
That is, the presence of the $s$ twist fields  gives
a big enough space $H^0(\oplus_{j=1..s}{{\cal O}_Z})$,
which helps to mod out
the
nontrivial regular sections of $V$. While this is not a
sufficient condition
for a well-behaved Landau-Ginzburg theory, presence of
a sufficient number of
local fermionic symmetries is required for stability of $V$.

For later reference, we note that the conditions $c_1(V)=c_1(TZ)=0$
and $c_2(V)=c_2(TZ)$ imply
\footnote{One can easily see the necessity of these conditions by
considering the corresponding Chern classes in the
Chow ring $A^*(\P)$. The sufficiency of these conditions on charges  
is not
assured unless $\P$ is itslef smooth, but we will neglect this subtlety
here.}
the following constraints on the multicharges :

\be
\sum_{\rho}{q_\rho} = -\sum_{\alpha = 1..t}{{\tilde q}_{0 \alpha }}
\ee
\be
\sum_{a=1..m}{{\tilde q}_a}=-q_0
\ee
\be
\sum_{a=1..m}{{\tilde q}^{k}_a{\tilde q}^{l}_a} -q_0^l
q_0^k =
\sum_{\rho}{q_\rho^{k} q_\rho^{l}}-\sum_{\alpha=1..t}{{\tilde  
q}^k_{0\alpha}
{\tilde q}_{0\alpha}^l}
\ee

Moreover, invariance of the action under the local
fermionic symmetries requires :
\be
\label{EF_constraint}
\sum_{a=1..m}{E_a^{(j)}F_a}=-\sum_{i=1..t}{E_{0i}^{(j)}G_i}
\ee

\section{Spectral Covers for $(0,2)$ Linear Sigma Models}

In this section we give a method for computing the spectral cover of the 
bundle $V$ in the case of a heterotic compactification on an elliptically 
fibered $n$-fold $Z$ which admits a $(0,2)$ linear sigma model description. 
No detailed understanding of subsection 4.1 is needed for 
reading the rest of this paper. The reader may consult only the summary of 
our method, given in subsection 4.2. 
Before giving the derivation, let us mention 
a few salient points:

(1)We will assume that the restriction $V_E$ 
of the bundle $V$ to the generic 
elliptic fibre $E$ of $Z$ is semistable. This is the main requirement for 
having a spectral cover description of $V$, as explained in \cite{FMW}.

(2)For simplicity, we will further assume that, for a 
generic $E$,  $V_E$ decomposes as a direct sum of line bundles. 
As explained in the appendix, this need not be the case, even if $V_E$ is 
semistable. 
However, a generic stable bundle $V$ over $Z$ (of trivial determinant) will 
have this property. Therefore, the simplified analysis we will present holds 
for the generic case.

(3)To avoid overwhelming technicalities, we will not explain here how one can 
actually test whether the restriction $V_E$ of a given $V$ to a generic 
elliptic fibre $E$ is semistable or not; neither will we explain how to  
test whether $V_E$ indeed decomposes in a direct sum of line bundles.

A significant number of bundles produced by $(0,2)$ model constructions fail 
to satisfy (1) or (2) or both. 
Therefore, far from being pedantic, a criterion for 
identifiying and dealing with such cases is a necessity once one 
starts exploring the wealth of available $(0,2)$ model constructions. 
While a bundle which 
violates (1) does not admit a spectral cover description, a bundle which 
violates (2) but obeys (1) is tractable. To study bundles of the latter type, 
one has to answer the questions raised at (3) and generalize 
the algorithm of this section to cases when $V_E$ does not split as a direct 
sum of line bundles. 
This can in fact be achieved by an extension of the 
methods presented here, as is discussed in detail in \cite{splitting_type}, 
to which we refer the interested reader.

\subsection{Derivation of the method}

\

\noindent {\bf Assumptions and notation} 

\

We will  be interested in the case when $Z$ is smooth, elliptically
fibered via a map $\pi_H : Z \rightarrow B_H$, and
has a nonsingular base $B_H$.
We will denote by $W_S$ the restriction of a vector bundle
$W \rightarrow Z$ to a sub{\em manifold} $S$ of $Z$.

Fix a point $b$ in the base $B_H$ and consider the elliptic fibre
$\pi^{-1}(b):=E$ above this point. Throughout our discussion we will keep
$b$ fixed,  treating it as a parameter. We assume that the  
fibration $\pi_H$
has a section $\sigma$. $\sigma(B_H)$ defines an effective divisor
in $Z$. We denote by $O_Z(\sigma)$ the associated line bundle on $Z$.
We assume that the restriction $V_E$ of $V$ is semistable and of trivial 
determinant for the generic elliptic fibre $E$ of $Z$; then
$V$ admits a spectral cover description as explained
in \cite{FMW,BJPS}.
Note that this is a nontrivial condition for $V$ and is the
main assumption underlying much of the work of \cite{FMW,BJPS}.
In particular, it is the only case for which a concrete proposal for the map
relating heterotic bundle data to $F$-theory data has been conjectured
\cite{FMW}.

Moreover, it can be argued \cite{FMW,BJPS}
that a `generic' stable bundle over $Z$ has the property that its
restriction to the generic elliptic fibre $E$ of $\pi_H$ is a fully split
semistable vector bundle. Here by fully split we mean that $V|_E$
splits as a direct sum of holomorphic line bundles
\footnote{This is somewhat
contrary to the usual mathematical convention in which splitting is  
considered
 for the associated $SU(r)$ principal bundle and is thus equivalent  
to the
existence of a filtration by holomorphic subbundles of consecutive  
dimensions.
Since such a filtration is automatic for degree zero semistable  
vector bundles
over (smooth) elliptic curves, this more classical terminology carries no
interesting information in our case.}. While this later assumption is not 
essential for the validity of our method of computing the cover 
\cite{splitting_type}, we will restrict the presentation of this section to 
bundles satisfying this additional condition. It should be noted, however, 
that this condition indeed fails to hold in a number of models that we 
were able to construct. Most models we include in the next section have been 
chosen to satisfy this assumption (this was tested by the 
methods of \cite{splitting_type}). 

\

\noindent {\bf Identifying the spectral cover}

\

Assuming that $b$ is a generic point in $B_H$ it folows that

\begin{equation}
V_E:=V|_E=\oplus_{i=1}^{r}{L_i}
\end{equation}
where $L_i$ are degree zero holomorphic line bundles over $E$.
Any such line bundle can
be represented as $L_i = O(q_i-p)$ where we choose the distinguished point 
$p$ of $E$ to be given by $p=\sigma(b) \in E$.
Now let $V'_E : = V_E \otimes O(p) $. $V'_E$ is a degree $r$
semistable vector bundle over $E$ which splits as

\begin{equation}
V'_E=\oplus_{j=1}^r{O_E(q_j)}
\end{equation}
The spectral cover associated to the bundle $V$ is essentially
the collection of
points $q_j$ over every point $b$ in the base. Therefore,
we are interested in concretely identifying the points $q_j$ starting
from our monad data.

\

\noindent {\bf Determining $q_j$ from sections of $V'_E$ } 

\

By the Riemann-Roch theorem we have $h^0(O(q_j))=1$,  
$h^1(O(q_j))=0$, which
shows that, under our assumptions for $V$ and $b$, $H^0(V'_E)$ is
$r$-dimensional while $H^1(V'_E)=0$.
Let $s_j \in H^0(O_E(q_j))$ be a holomorphic section with 
associated divisor $(s_j) = q_j$. 
Then  $s_j$ has a simple zero at the point $q_j$ and no other zeroes or 
poles.  Since  
dim$_{\C}H^0(O_E(q_j))=1$,
we have $H^0(O_E(q_j)) = <s_j>$ (the linear span of $s_j$).
(This implies in particular that each $s_j$ is unique up to  
multiplication by
a constant complex scalar).  It follows that $(s_1 ... s_r)$ form a  
basis of $H^0(V'_E)$.  Unfortunately, it is hard to  
identify such
sections of $V'_E$ directly.
Nevertheless, one can use any system of generators $(u_1 ... u_N)$ of the
vector space $H^0(V'_E)$ in order to determine the points $q_j$.  
For this, note that there must exist
a {\em constant} matrix $A$ with the property :

\begin{equation}
\label{A-relation}
(u_1 ... u_N)^t = A (s_1 ... s_r)^t
\end{equation}

Since 
$u_1 .. u_N$ generate $H^0(V'_E)$, we have rank$A$ $= r$. As
(\ref{A-relation}) is a functional equation, we can evaluate it at any
point $e\in E$ to obtain :

\begin{equation}
\label{A-relation_local}
(u_1(e) ... u_N(e))^t= A (s_1(e) ... s_r(e))^t
\end{equation}

Since rank$A = r$, it easy to see that dim$_{\C}<u_1(e) ... u_N(e)>$ =
dim$_{\C}<s_1(e) ... s_r(e)>$ for all $ e \in E$, where $< ... >$  
denotes the
linear span of the corresponding set of vectors
\footnote{Indeed, let  $(\epsilon_1(e) ... \epsilon_d(e))$ be a basis of
 $<s_1(e) ... s_r(e)>$, where
$d=$dim$_{\C}<s_1(e) ... s_r(e)>$. Then $(s_1(e) ... s_r(e))^t=
B(e)(\epsilon_1(e) ... \epsilon_d(e))^t$, with $B(e)$ a matrix 
of maximal rank. It follows that :\\
$ (u_1(e) ... u_N(e))^t= AB(e)(\epsilon_1(e) ... \epsilon_d(e))^t$.
As $A$ and $B$ have maximal rank, the associated linear operators
${\hat A } \in L(\C^r,\C^n)$ and ${\hat B} \in L(\C^d,\C^N)$ are
injective, and so is their composition
${\hat A} {\hat B}=\hat{AB}$, which implies that
$AB$ also has maximal rank, equal to $r$. Therefore
dim$_{\C}<u_1(e) ... u_N(e)>={\rm rank}(AB)=d=$dim$_{\C}<s_1(e) ...  
s_r(e)>$.}.
Now, it is clear
that dim$_{\C}<s_1(e) ... s_r(e)>$ will decrease in discrete jumps  
precisely
for $e \in \{q_1 ... q_r\}$ and this criterion can be used to
find the points $q_i$. It follows that the (multi)set ${q_1 ... q_r}$ can
be identified once one posesses any explicit system of generators
$(u_1 ... u_r)$ of $H^0(V'_E)$. Note that the points $q_i$ need not  
all be
distinct. If some subgroups of them coincide (say, $q_1=q_2= .. =  
q_{l_1} ,
q_{l_1+1}=..=q_{l_1+l_2},...,q_{l_1+..+l_{t-1}}= ... =q_{l_1+...+l_t}$,
with $l_1+...+l_t=r$) then dim$_{\C}<\sigma(e)...\sigma_r(e)>$ will
decrease by $l_j$ at the point $q_{l_{j-1}+1}=...=q_{l_{j-1}+l_j}$
and we can still use the information above to completely specify the
multiset $q_1 ... q_r$.
The calculation of the spectral cover is thereby reduced to
obtaining a system of generators of $H^0(V'_E)$ from our monad data,
and this is a simple exercise in sheaf cohomology, which we carry out next. 

\

\noindent {\bf Obtaining a generating set of sections from the monad 
description}

\

We start from the definition (\ref{monad}) of $V$, which we  
restrict to $E$
to obtain the following monad, whose cohomology defines the restriction
$V_E$
 of $V$ to $E$:
\be
0 \longrightarrow \oplus_{j=1}^{s}{O_E}
\stackrel{f}{\longrightarrow} \oplus_{a=1}^{m}{O_E(\tilde{q_a})}
\stackrel{g}{\longrightarrow} O_E(-q_0)\longrightarrow 0
\ee
Now twist the restricted monad by the line bundle $O_E(p)$ to obtain:

\be
0 \longrightarrow \oplus_{j=1}^{s}{O_E(p)}
\stackrel{f}{\longrightarrow} \oplus_{a=1}^{m}{O'_E(\tilde{q_a})}
\stackrel{g}{\longrightarrow} O'_E(-q_0)\longrightarrow 0
\ee
where we defined ${\cal E'}_E:={\cal E}_E\otimes O_E(p),
O'_E({\tilde q_a}):=O_E({\tilde q_a})\otimes
O_E(p)$ and $O'_E(-q_0):=O_E(-q_0)\otimes O_E(p)$.

It is easy to see that the cohomology of this  monad defines
a vector bundle which is canonically isomorphic to the twisted bundle
$V'_E:=V_E\otimes O(p)$.
We now rewrite this as a pair of exact sequences:

\begin{equation} 0 \longrightarrow \oplus_{j=1..s}O_E(p)
\stackrel{f}{\longrightarrow} \oplus_{a=1..m}O'_E({\tilde q_a})
\stackrel{p}{\longrightarrow} {\cal E'}\longrightarrow 0
\end{equation}

\begin{equation} 0 \longrightarrow V' \stackrel{j}{\longrightarrow}  
{\cal E'}
\stackrel{{\tilde g}}{\longrightarrow} O'(-q_0)\longrightarrow 0
\end{equation}
where ${\cal E}:={\rm coker}(f)$ and ${\tilde g}$ is the map canonically
induced by
$g$ via descent. Note that ${\tilde g}$ is the unique map making the
following diagram commute:

\be
\label{triangle}
\begin{array}{ccc} \oplus_{a=1..m}O'_E({\tilde q_a}) &\ &\ \\
				 p \downarrow & \bigcirc &\searrow {g} \\
		{\cal E}'_E & \stackrel{\tilde g}{\longrightarrow}  
&O'(-q_0)
\end{array}
\ee
Here $j$, $p$ act as the natural injection, respectively surjection  
on each
fiber. More precisely, for any $e \in E$, the action  of $p$ at $e$  
is given by
the canonical surjection:
\be
p_e : \oplus_{a=1..m}{O'_{E,e}({\tilde q_a})}\longrightarrow {\cal  
E'}_{E,e}
:=\oplus_{a=1..m}{O'_{E,e}({\tilde  
q_a})}/f_e(\oplus_{j=1..s}{O_{E,e}(p)}).
\ee
By using $H^1(O_E(p))=0$ and $H^1(V'_E)=0$ we see that the associated
cohomology long sequences collapse to the pair of short exact sequences

\be
\label{seq_1}
0 \longrightarrow \oplus_{j=1..s}{H^0(O_E(p))}
\stackrel{f_*}{\longrightarrow}\oplus_{a=1..m}{H^0(O'_E({\tilde q_a}))}
\stackrel{p_*}{\longrightarrow} H^0({\cal E'}_E) \longrightarrow 0
\ee

\be
\label{seq_2}
0 \longrightarrow H^0(V'_E)\stackrel{j_*}{\longrightarrow}  
H^0({\cal E'}_E)
\stackrel{{\tilde g}_*}{\longrightarrow}H^0(O'_E(-q_0))  
\longrightarrow 0.
\ee
where $f_*$,${\tilde g}_*$,$p_*$,$j_*$ denote the maps induced in degree
zero cohomology.
These maps act fiberwise on sections in the natural manner,
for example
$f_*(\eta)(e)=f_e(\eta(e))$, for any $\eta \in
\oplus_{j=1..s}{H^0(O_E(p))}$ and all $e \in E $. In particular, for all
$s \in \oplus_{a=1..m}{H^0(O'_E({\tilde q_a}))}$ and all $e \in E$  
we have:
\be
p_*(s)(e)=p_e(s(e))=s(e) { \ \rm mod \ }
\{ f_e(\oplus_{j=1..s}{O_{E,e}(p)})\} = s(e) +
f_e(\oplus_{j=1..s}{O_{E,e}(p)}).
\ee
The diagram (\ref{triangle}) gives
${\tilde g}_*\circ p_* = g_*$.
Together with the above sequences, this gives
the diagram :

\be
\label{diagram}
\begin{array}{ccccccccc}
\ &  \ &  \ & \ & 		0  &                   \ &  \ &   \  
&  \ \\
\ &  \ &  \ & \ &	   \downarrow  &	       \ & \ & \ &  \ \\
\ &  \ & \ & \ &	   \oplus_{j=1..s}{H^0(O_E(p))}	& \ & \ &   
\ &  \ \\
\ &  \ & \ & \ &	   f_*\downarrow  &           \ &   \ &   \  
 &  \ \\
\ &  \ & \ & \ &   \oplus_{a=1..m}{H^0(O'_E({\tilde q_a}))} \ & \ &  
\ & \ \\
\ &  \ & \ & \ &	   p_*\downarrow  &    \bigcirc & \searrow  
g_* &  \ & \ \\
0 &\longrightarrow &H^0(V'_E)&\stackrel{j_*}{\longrightarrow}& H^0({\cal
E'}_E)&
\stackrel{{\tilde g}_*}{\longrightarrow}&H^0(O'_E(-q_0))  
&\longrightarrow 0 \\
\ &  \ & \ & \ &           \downarrow			\ &  \ &    
\ &  \ \\
\ &  \ & \ & \ &		0			\ &  \ &   \ &  \
\end{array}
\ee
From (\ref{diagram}) one easily deduces that
\be
\label{final}
H^0(V'_E)=p_*({\rm ker}\{ \oplus_{a=1..m}{H^0(O'_E({\tilde q}_a))}
\stackrel{g_*}{\rightarrow} H^0(O_E'(-q_0)) \})
\ee
Moreover, since $p_*$ is surjective, the rank theorem for $p_*$ gives:
dim$_{\C} \oplus_{a=1..m}{H^0(O'_E({\tilde q}_a))}\\
=$ dim$_{\C}H^0({\cal E'}_E)+s$,
where we used ker$p_*=$im$f_*$ and injectivity of $f_*$ to deduce that
def$p_*=s$. Taking dimensions in the exact row of the diagram gives
dim$_{\C}H^0(O'_E({\tilde q}_a))=r+$dim$_{\C}H^0(O_E'(-q_0))$.
These relations combined imply :
\be
{\rm dim}_{\C} \oplus_{a=1..m}{H^0(O'_E({\tilde q}_a))}={\rm
dim}_{\C}H^0(O_E'(-q_0))+r+s
\ee
On the other hand, applying the rank 
theorem to the map $p_*|_{kerg_*}$ and noting from the diagram that
ker$p_* \subset {\rm kerg}_* ,{\rm im}(p_*|_{kerg_*})= p_*({\rm ker}g_*)=
H^0(V'_E)$ and ker$p_*=$im$f_*$ gives :
\be
N:={\rm dim}_{\C}{\rm ker}g_*=r+s
\ee
Note that none of these results requires that $V'_E$ is fully  
split, so they
will hold for any $V'_E$ semistable and of degree zero.

Now suppose that we are able to obtain a basis $(v_1 ... v_N)$
of ker $g_*$. $p_*$ being surjective, we are then assured that the sections
$u_1:=p_*(v_1) ... u_N:=p_*(v_N)$ form a system of generators of  
$H^0(V'_E)$
over ${\C}$. Now let $w_1(e) .. w_s(e)$ be a ${\C}$-basis
of $f_e(\oplus_{j=1..s}{O_{E,e}(p)})$. Then it is easy to see that  
\footnote{
Indeed, let $F_e:=<w_1(e) .. w_s(e)> =  
f_e(\oplus_{j=1..s}{O_{E,e}(p)})$ and let 
$S_e:=<v_1(e) ... v_N(e)>$. Then we have 
$S_e+F_e=<v_1(e) ... v_N(e) ,w_1(e) ... w_N(e)> $ and
$<u_1(e) ... u_N(e)> = p_e(<v_1(e) ... v_N(e)>)=p_e(S_e)$. Therefore :\\
dim$_{\C}<u_1(e) .. u_N(e)> =$  
dim$_{\C}p_e(S_e)=$dim$_{\C}S_e-$dim$_{\C}({\rm
ker}p_e\cap S_e)=
$dim$_{\C}S_e-$dim$_{\C}(F_e\cap S_e)\\
=$dim$_{\C}S_e -$dim$_{\C}F_e-$dim$_{\C}S_e+$dim$_{\C}(F_e+S_e)=
$dim$_{\C}<v_1(e) ... v_N(e) ,w_1(e) ... w_N(e)>-s $.}:

\be
\label{lift}
{\rm dim}_{\C}<u_1(e) ... u_N(e)>={\rm dim}_{\C}<v_1(e) ...  
v_N(e),w_1(e) ...
w_s(e)> -s
\ee
and we can identify $q_i$ by looking for jumps of
dim$_{\C}<v_1(e) ... v_N(e),w_1(e) ... w_s(e)>$.

\

\noindent {\bf A matrix formulation} 

\

A practical way to compute dim$_{\C}<v_1(e) ... v_N(e),w_1(e) ... w_s(e)>$ 
is to note that all of our vectors belong to
$\oplus_{a=1..m}{O'_{E,e}({\tilde q}_a)}$, so we may expand them in their
components (denoted by upperscript $a$) along the {\em one dimensional}
vector spaces $O'_{E,e}({\tilde q}_a)$.
Then dim$_{\C}<v_1(e) ... v_N(e),w_1(e) ... w_s(e)>$ can be obtained as
the rank of the matrix :
\be
{\tilde S}(e) :=\left( \begin{array}{cccccc}
	v_1^{(1)}(e) & ... & v_N^{(1)}(e) & , w_1^{(1)}(e) & ... &   
w_s^{(1)}(e)	\\
	... & ... & ...   & ... & ...  &  ....	                     
    \\
	v_1^{(m)}(e) & ... & v_N^{(m)}(e) & , w_1^{(m)}(e) & ... &   
w_s^{(m)}(e)
\end{array} \right)
\ee
where we define the rank by imploying the minors in the usual
way, except that in the definition of determinants we replace
multiplication by tensor product. (It is immediate
that this definition has the standard properties due to the fact that all
components of our vectors live in one-dimensional subspaces, so  
that one can
choose bases of these subspaces to reduce to the usual case of  
matrices over
the field of complex numbers).

Our task will  be finished if we can find a family $(w_1(e) ... w_s(e))$
of bases of the spaces  $f_e(\oplus_{j=1..s}{O_{E,e}(p)})$ for all  
$e \in E$.
It turns out that this task can be disposed of in the following manner.
Consider a local section $\eta \in H^0(U,O_E(p))$ of $O_E(p)$
above  an open neighborhood $U$ of $p$. We can always choose $\eta$  
such that
$\eta(e) \neq 0$. As the map
$f_e:\oplus_{j=1 .. s}{O_{E,e}(p)} \rightarrow \oplus_{a=1 .. m}
{O_{E,e}{O'({\tilde q_a})}}$ is injective for all $e \in E$, we are then
assured that
$(w_1(e):=f_e(\eta(e),...,0),...,w_s(e):=
f_e(0,...,\eta(e)))$ forms a $\C$-basis of  
$f_e(\oplus_{j=1..s}{O_{E,e}(p)})$.
In our case :
\be
w_j(e)=(E_1^{(j)}(e)\otimes \eta(e) ... E_m^{(j)}(e) \otimes \eta(e))
\ee
which shows that $\eta(e)$ will factor out of all minor  
determinants involved
in the computation of rank${\tilde S}(e)$. Thus, we have  
rank${\tilde S}(e)=$rank$S(e)$, where :
\\
\\
\be
\label{S}
S(e):=\left( \begin{array}{cccccc}
	v_1^{(1)}(e) & ... & v_N^{(1)}(e) & , E_1^{(1)}(e) & ... &   
E_1^{(s)}(e) \\
	...       & ... & ...       & , ...       & ... &  ....      \\

	v_1^{(m)}(e) & ... & v_N^{(m)}(e) & , E_m^{(1)}(e) & ... &   
E_m^{(s)}(e)
\end{array} \right)
\ee
\\
\\
and to find the points $q_1 ... q_r$ it suffices to look for jumps in
rank$S(e)$ as $e$ varies in $E$.

As is obvious from (\ref{A-relation_local}) and (\ref{lift}), we have
rank$S(e)=r+s$ for $e \in E-\{q_j|j=1..r\}$. To understand how this  
occurs,
note that, in virtue of (\ref{final}) and of the fact that
dim$_{\C}{\rm ker}p_*=
$dim$_{\C}({\rm ker}g_*)-$dim$_{\C}H^0(V'_E)=N-r=s$, there must be
$s$ independent linear combinations of $v_1 ... v_N$, with {\em constant}
coefficients, which give sections of $f(\oplus_{j=1..s}{O_E(p)})$.
Any such combination is of the form:

\be
\label{dependence}
\sum_{l=1..N}{\alpha_l v_l(e)}=\sum_{j=1..s}{\beta_j(e)\nu_j(e)}
\ee
where $\nu_1(e):=f_*(\nu,0...0), ... , \nu_s(e):=f_*(0,0,...\nu)$
and $\nu \in H^0(O_E(p))-\{ 0 \}$ is arbitrarily chosen.
(Indeed, since $f_{*,e}$
is injective and $\nu(e) \neq 0 $ for all $e \in E - \{ p \}$, we  
see that
$(\nu_1(e) ... \nu_s(e))$ form a basis of
$f_e(\oplus_{j=1..s}{O_{E,e}(p)})$ for all $e \in E-\{ p \}$. Then  
a simple
argument shows that (\ref{dependence}) will hold for all $ e \in E$ if we
allow $\beta_j(e)$ to be rational functions on $E$ with at most a pole of
order 1 at $p$ and no other poles.) Here $\alpha_l$ are {\em constant}
on $E$. Since

\be
\nu_j(e)=(E_1^{(j)}(e)\otimes \nu(e) ... E_m^{(j)}(e) \otimes \nu(e)),
\ee
we can rewrite (\ref{dependence}) as :
\be
\sum_{l=1..N}{\alpha_l v_l(e)}=\sum_{j=1..s}{\gamma_j(e)\otimes  
E^{(j)}(e)},
\ee
where $E^{(j)}(e)$ denotes the column vector
$(E_1^{(j)}(e) ... E_m^{(j)}(e))$ and where
$\gamma_j(e):=\beta_j(e)\otimes \nu(e)$ are {\em regular} sections of
$O_E(p)$. This gives $s$ linear dependences among the columns of $S(e)$,
which involve constant coefficients for $v_l(e)$. Therefore, we can use
these dependencies to {\em globally} eliminate $s$ of the sections
$v_1 .. v_N$
over $E$. One we did that, we are left with a subset $v_{i_1} ...  
v_{i_r}$
which descends to a basis $s_1:=p_*(v_{i_1}) ... s_r:=p_*(v_{i_r})$  
of the
$\C$-vector space $H^0(V'_E)$ and the matrix
$S(e)$ can be reduced to the $(r+s+1)\times(r+s)$ matrix :
\\
\\
\be
\label{S_0-matrix}
S_0(e):=\left( \begin{array}{cccccc}v_{i_1}^{(1)}(e) & ... &  
v_{i_r}^{(1)}(e) &
, E_1^{(1)}(e) & ... &  E_1^{(s)}(e) \\
	...       & ... & ...       & , ...       & ... &  ....      \\

	v_{i_1}^{(m)}(e) & ... & v_{i_r}^{(m)}(e) & , E_m^{(1)}(e)  
& ... &
E_m^{(s)}(e)
\end{array} \right)
\ee
\\
\\
such that rank$S(e)=$ rank$S_0(e)$ ,$\forall e \in E$.
Finding such dependencies can easily be achieved by looking for constant
solutions $\alpha_1 ... \alpha_N$ of the equation :

\be
\label{basis}
{\rm det}
\left( \begin{array}{ccccc}
   \sum_{l=1..N}{\alpha_l v_l^{(1)}(e)}\ , & E_1^{(1)}(e) & ... &  
E_1^{(s)}(e)
\\
   ... & ... & ... & ... \\
   \sum_{l=1..N}{\alpha_l v_l^{(m)}(e)}\ , & E_m^{(1)}(e) & ... &  
E_m^{(s)}(e)
\end{array} \right)
=0
\ee
Then one finds $v_{i_1} ... v_{i_r}$ in the obvious way.

Now let $\Delta_a(e)\in H^0(O_E(r- q_0 -{\tilde q_a}))$ $(a=1..m)$ be the
$(r+s) \times (r+s)$ -minors of $S_0(e)$ obtained by deleting the line
$(a)$ of $S_0(e)$).
We see that rank$S_0(e)=r+s-1$ iff $\Delta_a(e)=0 $ ,
$\forall a =1 .. m$.
We claim that $\Delta_a(e)=G_a(e)\otimes \phi(e)$
where $\phi \in H^0(O_E(r))$
\footnote{The easiest way to see this is to note
that for a fixed $e \in E$, we can choose local bases of the fibres
$O'_{E,e}({\tilde q}_a)$ and thus reduce to the case of complex numbers.
Then we can treat the columns $x_1 ... x_{r+s}$ of $S_0(e)$ as vectors
in the vector space $\C^m=\C^{r+s+1}$. Now we can consider the natural
nondegenerate {\em bilinear} form on $\C^m$ given by
$(x,y) = \sum_{a=1..m}{x_a y_a}$ and the natural mixed product  
given by the
$m$-linear form $(x^1 ...  
x^m):=\epsilon_{i_1...i_m}x^1_{i_1}...x^m_{i_m}$.
Associated to these objects there is a natural $(m-1)$ -linear vector
product (denoted by $\times$)
on $m-1$ ordered sets of vectors.
Now consider the vector $\Delta:=x_1\times ...
\times x_{r+s}$, whose components coincide with $\Delta_i(e)$.
Because all $x_i$ ($i=1..r+s$) lies in ker$(g_e)$, one sees immediately
that all $x_i$ are $( . , . )$ - orthogonal on the vector
$G:=(G_1(e) ... G_m(e))$.
Therefore, their vector product $\Delta$ is colinear to $G$, and  
since $G$ is
nonzero (as the map $g$ is transverse), we can write
$\Delta_a(e)=G_a(e)\otimes \phi(e)$. Because $\Delta_a$ and $G_a$ are
regular
and because $G_a$ are never all zero on $E$, this implies that  
$\phi$ is a
regular section of a line bundle on $E$, whose degree is then fixed  
to be $r$
by the degrees of $\Delta_a$ and $G_a$.}.
As $G_a$ never vanish simultaneously on $E$, we see that rank$S(e)<r+s$
iff $\phi(e)=0$.
We deduce that the {\em set} $\{ q_1 ... q_r \}$ coincides
with the support of the zero divisor $(\phi)$.

In conclusion, we can obtain the points $q_1 ... q_r $ ---
and thus the spectral cover --- provided that we can obtain a
basis of
 $H^0\left[{\rm ker} \{ \oplus_{a=1..m}{H^0(O'_E({\tilde q}_a))
\stackrel{g_*}{\rightarrow}H^0(O_E'(-q_0))} \}\right]$.
Since one knows the concrete form of $g_*$, this is straightforward to
achieve once one is able to obtain bases for each of the  
vector spaces $H^0(O'_E({\tilde q}_a))$, respectively $H^0(O_E'(-q_0))$. 
To achieve this in practice, one has to start with a concrete realization of 
$Z$. 

\

\noindent {\bf The construction of $Z$} 

\

In order to build examples with an `obvious' fibration structure,  
one can use
the methods of toric geometry. The underlying idea is to build $Z$ inside
a toric variety $\P$ which is itself fibered over the base $B_H$ with
fibre
a compact toric variety ${\P_f}$. In this case,
one also takes
the base $B_H$ to be toric. Then $Z$ is realized inside ${\P}$ as a
`generalized Weierstrass model' in which case the elliptic fibre is
visible by simply freezing the variables
corresponding to a point in $B_H$. The fibres of $\pi_H$ will then
naturally sit inside the fibres ${\P_f}$ of the fibration of the toric
ambient space.
The structure of the models involved is very simple
and will become apparent in the examples we give in the next section.
For all of those examples  we will have ${\P_f}=\W\P_{2,3,1}$ and  
$Z$ will be a
hypersurface in ${\P}$.

\

\noindent {\bf Descent from $\P_f$} 

\

If $L$ is a bundle over $\P_f$, a 
basis  of sections for its restriction $L_E$ to $E$   
can usually be obtained by descending from the ambient space
${\P_f}$ of $E$. In favorable cases --- as when
${\P_f}=\W\P_{2,3,1}$ ---
one obtains that the restriction maps $H^0(L) \stackrel p
{\longrightarrow} H^0(L_E)$ for the bundles  involved in our 
construction are surjective. 
In this case, one has a commutative diagram with {\em surjective}  
vertical
maps:
\vskip.3in
$\begin{array}{ccc} \oplus_{a=1..m}{H^0(O'_{\P_f}(q_a))} &
	\stackrel{g_*}{\rightarrow} & H^0(O_{\P_f}'(-q_0))	\\ p\downarrow
	&				\bigcirc &	\downarrow p'	\\
	\oplus_{a=1..m}{H^0(O'_E(q_a))} & \stackrel{g_*}{\rightarrow} &
	H^0(O_E'(-q_0))
\end{array}$
\vskip.3in
\noindent
which gives:
\be
\ker\{\oplus_{a=1..m}{H^0(O'_E(q_a))}
\stackrel{g}{\rightarrow}
H^0(O_E'(-q_0))\}=p({\rm ker}\{\oplus_{a=1..m}{H^0(O'_{\P_f}(q_a))}
\stackrel{p' \circ g_*}{\rightarrow}H^0(O_E'(-q_0))\})\nonumber
\ee

\

\noindent {\bf Sections of line bundles over $\P_f$}

\

In the examples we consider in section 5, if $\P$ has homogeneous 
coordinates $x_1..x_n$, then the base $B_H$ is  
described
by some
subset of the homogeneous coordinates ($x_1 .. x_k$), say, so  
fixing $b$ is
equivalent to fixing these coordinates to some values $b_1 .. b_k$ (up to
the $(\C^*)^\rho$ action). Then ${\P_f}$ is a toric variety whose
homogeneous
coordinates can be naturally identified with the restricions of 
$x_{k+1} ... x_n$ to $\P_f$. Thus, the
restriction of a line bundle $O(s_1...s_n)$ over ${\P}$ to a fixed
${\P_f}$ fibre is
nothing but the line bundle $O(s_{k+1} ... s_k)$ over ${\P_f}$.

To compute a $\C$-basis of the space of  
sections of a given line bundle
$O(s_{k+1}...s_n)$ over ${\P_f}$, recall  that there exists a vector space 
isomorphism
between $H^0({\P_f},O(s_{k+1} ... s_n))$ and the subspace of
polynomials in
$\C[x_{k+1} .. x_n]$ which are $(\C^*)^{n-k}$-multihomogeneous of
multihomogeneity degree $(s_{k+1} ... s_n)$. This reduces the computation
of $H^0({\P_f},O(s_{k+1} ... s_n))$ to finding all monomials of  
approprate multidegree, which is an algorithmic problem easily tackled by a  
computer.

\subsection{Summary of the procedure }

Let $E$ be a smooth elliptic fiber of $Z$.
Assuming that $V_E$ is fully split and semistable,
the points where the spectral cover of $V$ intersects $E$ form
a multiset $q_1 ... q_r$. To determine this multiset consider the twisted
bundle $V'_E$ defined by the monad :

\be
0\longrightarrow \oplus_{j=1}^{s}{O_E(p)}
\label{summary_seq}
\stackrel{f}{\longrightarrow} \oplus_{a=1}^{m}{O'_E(\tilde{q_a})}
\stackrel{g}{\longrightarrow} O'_E(-q_0)\longrightarrow 0
\ee
Then compute a basis of sections $(v_1 ... v_N)$ ($N = r+s$) of the
$\C$-vector space:
\be
{\rm ker}\{ \oplus_{a=1..m}{H^0(O'_E({\tilde q}_a))
\stackrel{g_*}{\rightarrow}H^0(O_E'(-q_0))} \}
\ee
and form the matrix of sections $S$ given in (\ref{S}).
The  multiset $ q_1 ... q_r$ can be identified as follows.
The underlying set $\{q_j|j=1..r\}$ is the set
$Z(S):=\{t \in E | {\rm rank}S(t) < r+s\}$.
The multiplicity of any $t\in S$ in the
multiset $ q_1 ... q_r$ is given by $m_t:=r+s-{\rm rank}S(t)$.
That is, the desired
multiset is given by $\{m_t t |t \in S\}$.

Moreover, if one wishes to compute a basis $s_1 ... s_r$ of sections
of $V'_E$, then one can solve the equation (\ref{basis}) for {\em  
constants}
$\alpha_1 ... \alpha_N$. This will give a set of $s$ linear  
relations with
constant coefficients among $v_1 ... v_N$ (modulo the $E$'s). By  
eliminating
$s$ of the sections $v_1 ... v_N$ via these relations, one obtains  
a subset
of sections $v_{i_1} ... v_{i_r}$ of  
$\oplus_{a=1..m}{H^0(O'_E({\tilde q}_a))}$
whose natural descendants $s_1:=p_*(v_{i_1}) ... s_r:=p_*(v_{i_r})$  
form a
basis of the vector space $H^0(V'_E)$.
If one has already determined such a basis, then one can form the matrix
$S_0$ in equation (\ref{S_0-matrix}).

Now find all $(r+s)\times(r+s)$ - minors of $S$.
For each line $a \in \{ 1..m \}$ of $S$, there is exactly one
such minor $\Delta_a$, obtained by deleting the line $(a)$.
This gives a set of $m=r+s+1$ such minors,
labeled $\Delta_a$, $a=1..m$. Then there exists a regular section $\phi$
of $O_E(r)$ such that $\Delta_a=G_a\otimes \phi$ for all $a=1..m$.
The underlying set $\{ q_i | i=1..r \}$ of the multiset $q_1 ... q_r$
coincides with the support of the zero divisor of the section $\phi$.
If the support of this divisor is formed of less than $r$ points,then one
can deduce the multiplicity of each point $q \in {\rm supp}(\phi)$ in
the multiset $q_1 ... q_r$ by computing rank$S(q)$ at that point.

\section{Examples}

\subsection{Models on a $K3$ surface realized as a hypersurface in a 
resolution of $\W\P^{6,4,1,1}$}

\subsubsection{A model on the tangent bundle}

We first present a model on a $K3$ surface $Z$ realized as a degree $(12,6)$
hypersurface in a resolution $\P$ of $\W\P^{6,4,1,1}$. 
The bundle will simply
be its tangent bundle. In the language of $(0,2)$ linear sigma models
we have $6$ right-moving bosonic fields ($x_1,..x_5,p$)
and $6$ left moving fermionic fields ($\lambda_1,..,\lambda_5,\gamma$)
which are charged as follows under the $U(1)^2$ worldsheet gauge group :

$$
\hbox{
\vbox{\offinterlineskip \tabskip=0pt
\halign{
#&
\vrule height 10pt depth 5pt
\enskip\hfil$#$\hfil\enskip\vrule &
\enskip\hfil$#$\hfil\enskip\vrule &
\enskip\hfil$#$\hfil\enskip\vrule &
\enskip\hfil$#$\hfil\enskip\vrule &
\enskip\hfil$#$\hfil\enskip\vrule &
\enskip\hfil$#$\hfil\enskip\vrule &
\enskip\hfil$#$\hfil\enskip\vrule &
\enskip\hfil$#$\hfil\enskip\vrule \cr\tablerule&
x_1&
x_2&
x_3&
x_4&
x_5&
p\cr\tablerule&
6&4&1&1&0&-12\cr\tablerule&
3&2&0&0&1&-6\cr\tablerule
}}
\qquad
\qquad
\vbox{\offinterlineskip \tabskip=0pt
\halign{
#&
\vrule height 10pt depth 5pt
\enskip\hfil$#$\hfil\enskip\vrule &
\enskip\hfil$#$\hfil\enskip\vrule &
\enskip\hfil$#$\hfil\enskip\vrule &
\enskip\hfil$#$\hfil\enskip\vrule &
\enskip\hfil$#$\hfil\enskip\vrule &
\enskip\hfil$#$\hfil\enskip\vrule \cr\tablerule&
\lambda_1&
\lambda_2&
\lambda_3&
\lambda_4&
\lambda_5&
\gamma\cr\tablerule&
6&4&1&1&0&-12\cr\tablerule&
3&2&0&0&1&-6\cr\tablerule
}}}
$$
\def\nn{\nonumber}
Here each line of a table gives the charge under the corresponding $U(1)$ 
group.

We will choose the following $F$'s and $G$ as a definition of the complex
structure of the $K3$ surface and part of the definition of the bundle:

\bearray
G &= &x_1^2 + x_2^3 + (x_3^{12} + x_4^{12}) x_5^6 \nonumber \\
F_1 &=& 2 x_1  \nonumber \\
F_2 &=& 3 x_2^2	 \nn \\
F_3 &=&	 12 x_3^{11} x_5^6 \nn \\
F_4 &=&	 12 x_4^{11} x_5^6 \nn \\
F_5 &=& 6 (x_3^{12} + x_4^{12}) x_5^5
\eearray

We will also need to specify two sets of $E$'s, which are given as follows:
$$
\medskip\centerline{
\vbox{\offinterlineskip \tabskip=0pt
\halign{
#&
\vrule height 10pt depth 5pt
\enskip\hfil$#$\hfil\enskip\vrule &
\enskip\hfil$#$\hfil\enskip\vrule &
\enskip\hfil$#$\hfil\enskip\vrule &
\enskip\hfil$#$\hfil\enskip\vrule &
\enskip\hfil$#$\hfil\enskip\vrule &
\enskip\hfil$#$\hfil\enskip\vrule \cr\tablerule&
E_1&
E_2&
E_3&
E_4&
E_5&
E_{01}
\cr\tablerule&
0 & 0 & x_3 & x_4 & -2 x_5 & 0 \cr\tablerule &
3 x_1 & 2 x_2 & 0 & 0 & x_5 & -6 x_6 \cr\tablerule
}}}
$$
One easily checks that each set of $E$'s satisfies the condition
 $\sum E_i F_i + E_{01} G = 0$. The transversality conditions  
are also obeyed.

The fibration of $Z$ is obtained by viewing $x_3$, $x_4$ as the base 
parameters. Then the equation $G=0$ gives the elliptic fibre as 
a sextic curve inside of the
$WP_{3,2,1}$ fibre (spanned by the homogeneous
coordinates $x_1,x_2,x_5$) of the ambient toric space $\P$. The base of the 
fibration is a $\P^1$.
It is easy to see that the toric divisor $(x_5)$ cuts $Z$ along a section
$\sigma=Z \cap (x_5)$ of the elliptic fibration. Therefore, 
an elliptic fibre 
$E$ meets $\sigma$ in the point $p=E \cap (x_5)$ and we have 
$O_E(p)=O_E(x_5)$, where $O_E(x_5):=O(x_5)|_E=O(0,1)|_E:=O_E(0,1)$.

Given the above data, it is now easy to compute the spectral cover via
the technique introduced in section 4.
In the notation of (\ref{summary_seq}), the twisted bundle 
$V'_E=V_E\otimes O_E(p)$ is given by the complex :

\be
0\longrightarrow \oplus_{j=1}^{2}{O_E(p)}
\stackrel{f}{\longrightarrow} O_E(6,4) \oplus O_E(4,3) \oplus O_E(1,1) 
\oplus O_E(1,1) \oplus O_E(0,2) 
\stackrel{g}{\longrightarrow} O_E(12,6)\longrightarrow 0 
\ee

\noindent where we twisted the defining monad of $V_E$ 
by $O_E(x_5)=O_E(0,1)$.
The first step of the procedure involves finding the kernel of $g_*$.
This is spanned by the columns of the following matrix of sections 
\footnote{Here (and in all following examples) we decomposed each section
of ${\rm ker}g_* \subset \oplus_{a=1..m} {O'_E({\tilde q_a})}$ in its 
components
along $O'_E({\tilde q_a})$. The lines of the matrix are indexed by 
$a=1..m$ in this order, while each column gives the components of a section
of ${\rm kerg}_*$. The set of these columns forms a basis of ker$g_*$.}
:

\be
K:=\left [\begin {array}{cccc} {\frac {3\,x_{{1}}x_{{5}}}{
2}}&-{\frac {3\,{x_{{2}}}^{2}}{2}}&0&0
\\\noalign{\medskip}x_{{2}}x_{{5}}&x_{{1}}&0&0
\\\noalign{\medskip}{\frac {\left ({x_{{3}}}^{12}+{x_{{
4}}}^{12}\right )x_{{5}}}{4\,{x_{{3}}}^{11}}}&0&-{
\frac {{x_{{4}}}^{11}x_{{5}}}{{x_{{3}}}^{11}}}&{\frac {
\left (-\,{x_{{3}}}^{12}-\,{x_{{4}}}^{12}\right )x_{{
5}}}{2\,{x_{{3}}}^{11}}}\\\noalign{\medskip}0&0&x_{{5}}
&0\\\noalign{\medskip}0&0&0&{x_{{5}}}^{2}
\end {array}
\right ]
\ee

It is easy to see that columns 1 and 4 of the above matrix are linearly
dependent modulo constants multiplying columns 2,3 and the twisted $E$'s 
(i.e.
columns formed by multiplying the $E$'s by $x_5$), hence we may remove them
in constructing the matrix $S_0$ as explained above . The matrix
$S_0$ is thus constructed by appending the $E$'s to columns 2 and 3 of  
$K$ :
\be
S_0:=\left [\begin {array}{cccc} -{\frac {3\,{x_{{2}}}^{2}}{
2}}&0&0&3\,x_{{1}}\\\noalign{\medskip}x_{{1}}&0&0&2\,x_
{{2}}\\\noalign{\medskip}0&-{\frac {{x_{{4}}}^{11}x_{{5
}}}{{x_{{3}}}^{11}}}&x_{{3}}&0\\\noalign{\medskip}0&x_{
{5}}&x_{{4}}&0\\\noalign{\medskip}0&0&-2\,x_{{5}}&x_{{5
}}\end {array}\right ]
\ee
The spectral cover may then be found by dividing the $i$-th 4x4 minor of 
$S_0$ by $F_i$, leading to the equation 
$\left[(x_3^{12} + x_4^{12})^2\right] x_5^2=0$ (where a denominator given by 
$x_3^{11}$ was discarded).
Further analysis by the methods of \cite{splitting_type} 
shows that $V_E$ is actually $F_2$, 
an irreducible stable rank 2 bundle\footnote{This is denoted by 
$I_2$ in \cite{FMW_math}, but we use Atyiah's notation as in the 
appendix}. This agrees with the result of Corollary 6.7 of \cite{FMW_math}. 
Although this is not fully split, it will
be shown in \cite{splitting_type} that our method for computing the cover 
still goes through for such a case. 

\subsubsection{A generalization of the previous model}

Let us consider a generalization of the previous
model. Specifically, we still consider the tangent bundle of a K3 
surface realized in the same ambient space, but now we let the surface 
be given by a more general degree $(12,6)$ polynomial. The charges of all 
fields are as before, but now we take :
\be
G =x_1^2 + x_2^3 + x_2 x_5 ^4 f + g x_5^6 \nonumber 
\ee
\be
F_1 =\partial_1 G =2 x_1 \nonumber 
\ee
\be
F_2 =\partial_2 G = 3 x_2^2 + x_5 ^4 f  \nonumber
\ee
\be
F_3 =\partial_3 G =  x_2 x_5^4 \partial_3 f  +  \partial_3 g x_5^6 
\nonumber 
\ee
\be
F_4 =\partial_4 G = x_2 x_5^4 \partial_4 f   + \partial_4 g x_5^6 
\nonumber
\ee
\be
F_5 =\partial_5 G = 4  f  x_2 x_5 ^3+ 6 g x_5^5  \nonumber
\ee
where $f$ and $g$ are homogeneous polynomials in $(x_3, x_4)$  of  
multidegrees $(8,0)$, respectively $(12,0)$.
The elliptic fibration degenerates at the zeroes of the 
discriminant $\Delta =4f^3 + 27 g^2$ of the elliptic fibration. For generic
choices of $f,g$, the resulting $K3$ surface will be smooth. In particular, 
transversality of $F_1..F_5$ on $Z$ is assured in such a case.
Since we desire the tangent bundle of $Z$, the $E$'s are the same as before.
Again each set of $E$'s satisfies 
 $\sum E_i F_i + E_{01} G = 0$ and the fibration structure is similar to 
that of the previous example. The twisted bundle is given by the same 
complex.

The kernel of $g_*$ is spanned by the columns of the following matrix of 
sections :

\begin{equation}
K:=\left [\matrix{
0 & 0 & x_5 x_1 & -\frac{1}{2}(f x_5 ^4 + 3 x_2 ^2)\cr
-12 f x_5^3 \frac{\cal W}{\partial_4 \Delta}& -6 f x_5 ^3 x_3 \frac {\cal
W}{\partial_4 \Delta} & \frac{2}{3}x_5 x_2 + 2 f x_5 ^3 x_3 \frac{\cal
W}{\partial_4 \Delta} & x_1 \cr
x_5 & 0 & 0 & 0\cr
-x_5 \frac{ \partial_3 \Delta}{\partial_4 \Delta} & -12 x_5
\frac{\Delta}{\partial_4 \Delta}& 4 x_5 \frac{\Delta}{\partial_4  
\Delta} & 0\cr
9 x_2 \frac{\cal W}{\partial_4 \Delta} & x_5 ^2 + \frac{9}{2} x_2 x_3
\frac{\cal W}{\partial_4 \Delta} & -\frac{3}{2} x_2 x_3  \frac{\cal
W}{\partial_4 \Delta} & 0\cr
}\right ]
\end{equation}
where, to simplify formulae, we introduced the  Wronskian  ${\cal  
W}=\partial _3
g \partial _4 f - \partial _3 f \partial _4 g$.  It is easy to see  
that the
second and the third columns are
linear combinations of the first column and the twisted $E$'s.
The matrix $S_0$ is thus obtained by appending the $E$'s to columns one and  
four of $K$ :

\begin{equation}
S=\left [\matrix{
0 & 3x_1 & 0 & -\frac{1}{2}(f x_5 ^4 + 3 x_2 ^2)\cr
-12 f x_5^3 \frac{\cal W}{\partial_4 \Delta}& 2x_2 & 0 & x_1 \cr
x_5 & 0 & x_3 & 0\cr
-x_5 \frac{ \partial_3 \Delta}{\partial_4 \Delta} & 0& x_4& 0\cr
9 x_2 \frac{\cal W}{\partial_4 \Delta} & x_5 & -2 x_5 & 0\cr
}\right ]
\end{equation}

The spectral cover is again found by dividing the $i$-th 4x4 minor of
$S$ by $F_i$, leading to the equation $x_5 ^2 \Delta=0$ 
(where a denominator given by $\partial_4 \Delta$ was discarded). 
This is again degenerate. 

Note that the coefficient multiplying $x_5^2$ has zeroes along the 
discriminant locus of $Z$. This leads to the 
conclusion that the spectral cover must include the 24 degenerate elliptic 
fibers of $Z$. In particular, this is consistent with the formula 
(\ref{class3}) 
for the class of the spectral cover (if one valiantly generalizes it to 
this degenerate case). 

According to the interpretation proposed in 
\cite{BIKMSV}, such a cover would correspond to heterotic small instantons, 
but in our case the bundle $V$ is perfectly smooth. If the above statements 
are correct, it appears that the spectral bundle (or rather the spectral 
sheaf) must conspire with the degenerate elliptic fibres such that a 
(generalization of) the pushforward construction yields a well-defined 
{\em bundle} (i.e. a locally free sheaf). A more detailed discussion of 
this will be attempted elsewhere.

\subsubsection{A simple deformation of the tangent bundle}

As in the previous subsection, we consider a $K3$ surface realized as a 
degree $(12,6)$ hypersurface in a resolution of the weighted projective space 
$\W \P^{6,4,1,1}$. 
Let us require that the deformed bundle has the same set of $E$'s as the  
tangent bundle. This is essentially the simplest possible 
deformation.
Condition (\ref{EF_constraint}) severely constrains the allowed $F_j$. Namely,
one must have
\bearray
F_1 &=&2 x_1 -\frac{2}{3} h x_2 x_5 - k x_5^3\nonumber \\
F_2 &=& 3 x_2^2	+ x_5 ^4 (f -a) + h x_1 x_5  -  m x_2 x_5^2\nn \\
F_3 &=&  x_2 x_5^4 p_3  +  q_3 x_5^6  + r_3 x_1 x_5^3 + s_3 x_2^2 x_5^2\nn \\
F_4 &=&  x_2 x_5^4 p_4  + q_4 x_5^6  + r_4 x_1 x_5^3 + s_4 x_2^2 x_5^2\nn \\
F_5 &=&(4  f +2a ) x_2 x_5 ^3+ 6 g x_5^5  + 3 k x_1 x_5^2  + 2 m x_2^2 x_5~,
\eearray
where $p_3,p_4,q_3,q_4, h$ and $a$ are polynomials on the base of the  
appropriate degrees. (\ref{EF_constraint}) 
also implies two relations between them, namely
\begin{equation}
x_3 p_3 + x_4 p_4 = 8f + 4a ~~,~~~~x_3 q_3 + x_4 q_4 = 12 g~~,~~~~x_3 r_3 + 
x_4 r_4 = 6 k~~,~~~~x_3 s_3 + x_4 s_4 = 4 m
\end{equation}
Going through computations similar to the above,  
one obtains that the spectral curve is given by : 
\bearray
\Sigma &: & x_5^2 ~~(
576\,{f}^{3} + 3888\,{g}^{2}-32\,{h}^{2}m{f}^{2}+576\,hk{f}^{2}+
144\,{m}^{2}{f}^{2}-
288\,{h}^{2}fg-24\,{h}^{3}mkf\nn\\
&&-144\,a{m}^{2}f
+144\,hkaf+
120\,hk{m}^{2}f-432\,{a}^{2}f+108\,{h}^{2}{k}^{2}f+16\,m{h}^{2}af-72\,
{h}^{3}kg \nn\\
&&-144\,a{h}^{2}g
-96\,{h}^{2}{m}^{2}g 
+16\,{h}^{4}mg+1296\,amg-
216\,hmkg+1944\,{k}^{2}g+144\,{m}^{3}g+\nn\\
&&16\,m{h}^{2}{a}^{2}+36\,{m}^{3}
{k}^{2}+
324\,am{k}^{2}
-54\,h{k}^{3}m-48\,ahk{m}^{2}-144\,{a}^{3}+243\,{k}^{4}-\nn\\
&&72\,hk{a}^{2})=0
\eearray
(where a denominator dependent only on the base coordinates was discarded).
This is again a degenerate cover.

\subsection{Models over a Calabi-Yau threefold 
realized as a hypersurface in a resolution of 
$\W\P^{9,6,1,1,1}$}

\subsubsection{A model on the tangent bundle}

We first present a model on a Calabi-Yau threefold 
$Z$ realized as a degree $(18,6)$
hypersurface in a resolution $\P$ of $\W\P^{9,6,1,1,1}$. The bundle is the 
tangent bundle of $Z$.
The $(0,2)$ linear sigma model  has $7$ right-moving bosonic fields 
($x_1,..x_6,p$) and $7$ left moving fermionic fields 
($\lambda_1,..,\lambda_6,\gamma$)
with the $U(1)^2$ charges: 

$$\hbox{\hskip -2in\vbox{
\medskip\centerline{
\vbox{\offinterlineskip \tabskip=0pt
\halign{
#&
\vrule height 10pt depth 5pt
\enskip\hfil$#$\hfil\enskip\vrule &
\enskip\hfil$#$\hfil\enskip\vrule &
\enskip\hfil$#$\hfil\enskip\vrule &
\enskip\hfil$#$\hfil\enskip\vrule &
\enskip\hfil$#$\hfil\enskip\vrule &
\enskip\hfil$#$\hfil\enskip\vrule &
\enskip\hfil$#$\hfil\enskip\vrule &
\enskip\hfil$#$\hfil\enskip\vrule \cr\tablerule&
x_1&
x_2&
x_3&
x_4&
x_5&
x_6&
p\cr\tablerule&
9&6&1&1&1&0&-18\cr\tablerule&
3&2&0&0&0&1&-6\cr\tablerule
}}
}
}
\hskip -4in\vbox{\medskip\centerline{
\vbox{\offinterlineskip \tabskip=0pt
\halign{
#&
\vrule height 10pt depth 5pt
\enskip\hfil$#$\hfil\enskip\vrule &
\enskip\hfil$#$\hfil\enskip\vrule &
\enskip\hfil$#$\hfil\enskip\vrule &
\enskip\hfil$#$\hfil\enskip\vrule &
\enskip\hfil$#$\hfil\enskip\vrule &
\enskip\hfil$#$\hfil\enskip\vrule &
\enskip\hfil$#$\hfil\enskip\vrule \cr\tablerule&
\lambda_1&
\lambda_2&
\lambda_3&
\lambda_4&
\lambda_5&
\lambda_6&
\gamma\cr\tablerule&
9&6&1&1&1&0&-18\cr\tablerule&
3&2&0&0&0&1&-6\cr\tablerule
}}}
}}$$

We have :
\bearray
G &= &x_1^2 + x_2^3 + x_2 x_6 ^4 f + g x_6^6 \nonumber \\
F_1 &=&2 x_1  \nonumber \\
F_2 &=& 3 x_2^2	+ x_6 ^4 f  \nn \\
F_3 &=&\partial_3 f x_2 x_6^4 +  \partial_3 g x_6^6 \nn \\
F_4 &=&\partial_4 f x_2 x_6^4 +	 \partial_4 g x_6^6 \nn \\
F_5 &=&\partial_5 f x_2 x_6^4 +	 \partial_5 g x_6^6 \nn \\
F_6 &=&4  f  x_2 x_6 ^3+ 6 g x_6^5~,
\eearray
where $f$ and $g$ are homogeneous polynomials in
$(x_3, x_4, x_5)$  of degrees $(12,0,0)$ and $(18,0,0)$. The fermionic 
symmetries are specified by :
\begin{equation}
\label{ddd}
\medskip\centerline{
\vbox{\offinterlineskip \tabskip=0pt
\halign{
#&
\vrule height 10pt depth 5pt
\enskip\hfil$#$\hfil\enskip\vrule &
\enskip\hfil$#$\hfil\enskip\vrule &
\enskip\hfil$#$\hfil\enskip\vrule &
\enskip\hfil$#$\hfil\enskip\vrule &
\enskip\hfil$#$\hfil\enskip\vrule &
\enskip\hfil$#$\hfil\enskip\vrule &
\enskip\hfil$#$\hfil\enskip\vrule \cr\tablerule&
E_1&
E_2&
E_3&
E_4&
E_5&
E_6&
E_{01}
\cr\tablerule&
0 & 0 & x_3 & x_4 & x_5 & -3x_6 & 0 \cr\tablerule &
3x_1 & 2 x_2 & 0 & 0 & 0 & x_6& -6 \cr\tablerule
}}}
\end{equation}

The base of the elliptic fibration is a $\P^2$ parameterized by the  
homogeneous
coordinates $x_3,x_4,x_5$. The elliptic fibre is  a sextic in  
$WP_{3,2,1}$. As before, $\Delta =4f^3 + 27 g^2$ denotes the discriminant of 
the elliptic fibration.
The equation of the cover is 
$x_6 ^3 \Delta / \partial_4 \Delta=0$ (discarding a denominator given by 
$\partial_4 \Delta$), which is degenerate as well.

\subsubsection{Simple deformations of the tangent bundle}

As in the K3 case, one can consider the simple deformation 
of the tangent bundle which preserves the $E$'s.
This is given by :
\bearraynn
G &= &x_1^2 + x_2^3 + x_2 x_6 ^4 f + g x_6^6 \nonumber \\
F_1 &=&2 x_1 -\frac{2}{3}h x_2 x_6 \nonumber \\
F_2 &=& 3 x_2^2	+ x_6 ^4 (f-a) + h x_1 x_6  \nn \\
F_3 &=&p_3 x_2 x_6^4 +  q_3 x_6^6 \nn \\
F_4 &=&p_4 x_2 x_6^4 +	 q_4 x_6^6 \nn \\
F_5 &=&p_5 x_2 x_6^4 +	 q_6 x_6^6 \nn \\
F_6 &=&(4f + 2a)  x_2 x_6 ^3+ 6 g x_6^5~,
\eearraynn
The functions $p_i,~q_i$ satisfy the constraints
\begin{equation}
x_3 p_3 + x_4 p_4 +x_5 p_5= 12f + 6a ~~,~~~~x_3 q_3 + x_4 q_4+ x_5  
q_5 = 18 g
\end{equation}
The spectral cover is again degenerate and is given by:
\begin{equation}
\label{stcover}
 \Sigma: ~x_5 ^3 (4f^3 + 27 g^2 - 3 a^2 f - a^3 -h^2 a g - 2  h^2 f g )=0
\end{equation}
(where a denominator given by $p_6 q_4 - p_4 q_6$ was discarded).

Under the same assumptions as before, the spectral surface would 
consists of 3 copies of the section of $Z$ and a vertical  
component, which projects on the base $\P ^2$ as a curve $\gamma$
of degree 36, given by the zero locus of the polynomial 
$4f^3 + 27 g^2 - 3 a^2 f - a^3 -h^2 a g - 2 h^2 f g$.

So far we've obtained rather `boring' covers in the sense that they are
always merely multiple covers of the base of the heterotic manifold 
(modulo some extra elliptic fibers). 
One may wonder if a monad presentation can be found for bundles
with a nontrivial spectral cover. We will see that this is indeed
the case in the next paragraph; the model we will display is however {\em not} 
perturbatively well-defined as a heterotic string theory, due to 
problems at the LG point.

\subsubsection{A perturbatively unsound model on
the same hypersurface}

We will again consider a degree 18 hypersurface in $\W\P^{9,6,1,1,1}$.
In the previous example there are stringent constraints on the allowed 
deformations of the $F$'s because of the requirement of having 
transverse $E$'s. 
Here we will
relax that requirement and see where it leads us. In doing so, we  
obtain a
model whose spectral cover is nondegenerate but whose LG phase is not well
behaved. TIt is commonly believed that in such a case stability of 
$V$ over $Z$ fails.

Consider a general set of $F$'s and $G$:
\bearray
F_1 & = & 3\,{x_2^2} + n_2\,x_2\,{x_6^2} + n_4\,{x_6^4},\nn\\
F_2 & = & 2\,x_1 + p_1\,x_2\,x_6 + p_3\,{x_6^3},\nn\\
F_3 & = & f_4\,x_2\,{x_6^4} + f_6\,{x_6^6}, \nonumber \\
F_4 & = & g_4\,x_2\,{x_6^4} + g_6\,{x_6^6}, \nn \\
F_5 & = & h_4\,x_2\,{x_6^4} + h_6\,{x_6^6},\nn\\
F_6 & = & k_3\,x\,{x_6^3} + k_5\,{x_6^5},\nn\\
G & = & {x_1^2} + {x_2^3} + f\,x_2\,{x_6^4} + g\,{x_6^6}.
\label{unsound}
\eearray
The polynomials
$f_4$,$f_6$,
$g_4,~g_6,~h_4,~h_6,~n_2,~n_4,
{}~p_1,~p_3,~k_3,~k_5,~f$ and $g$
in (\ref{unsound}) are taken to have the appropriate degrees and 
to depend only on the coordinates on the base $B_H$.
This gives a transverse bundle of rank 5
over a smooth threefold, but
there are no longer suitable E's in general.

The spectral cover is given by:
\bearraynn
\Sigma : x_6^3 (a_0 x_6^2+ a_2 x_2)=0,
\eearraynn
where
$$ a_2= (-36\,g\,k_3+ 36\,f\,k_5
   - 12\,k_5\,n_4) +
    4\,k_5\,{n_2^2}-
     4\,k_3\,n_2\,n_4 -
     3\,k_5\,n_2\,{p_1^2} +
     3\,k_3\,n_4\,{p_1^2} +
     18\,k_5\,p_1\,p_3 -
     9\,k_3\,{p_3^2} ,
$$
$$ a_0= (36\,g\,k_5 +
    12\,f\,k_3\,n_4-4\,k_3\,{n_4^2} ) -
    12\,g\,k_3\,n_2 +
          4\,k_5\,n_2\,n_4 -
 3\,k_5\,n_4\,{p_1^2}+
     6\,k_3\,n_4\,p_1\,p_3+
     9\,k_5\,{p_3^2}-
     3\,k_3\,n_2\,{p_3^2} . $$
Note that this spectral cover is independent of 
$f_{4,6},~g_{4,6}$ and $h_{4,6}$.
For the tangent bundle $n_4 \sim f,~k_3 \sim f,~k_5 \sim g$
and all other coefficients (except $f_{4,6},~g_{4,6},~h_{4,6}$)
are equal to zero. In that case, the term in parenthesis in $a_2$   
vanishes and the term in parenthesis in $a_0$ 
coincide with the discriminant of the elliptic fibration.

We see that relaxing the condition of having fermionic symmetries
leads to a nondegenerate spectral cover, at the price of having a poorly 
defined $(0,2)$ linear sigma model. The situation is very similar to
what we found in the case of 6-dimensional compactifications.

\subsubsection{General deformations of the $E$'s}

In order to convert the perturbatively unsound model of the  
previous section
into a well defined one it is necessary to impose fermionic  
symmetries.  Here
we will assume that the fermionic symmetries are the deformations of
(\ref{ddd}), namely

$$\vcenter{
\halign{
#&
\vrule height 10pt depth 5pt
\enskip\hfil$#$\hfil\enskip\vrule &
\enskip\hfil$#$\hfil\enskip\vrule &
\enskip\hfil$#$\hfil\enskip\vrule &
\enskip\hfil$#$\hfil\enskip\vrule &
\enskip\hfil$#$\hfil\enskip\vrule &
\enskip\hfil$#$\hfil\enskip\vrule &
\enskip\hfil$#$\hfil\enskip\vrule \cr\tablerule&
E_1&
E_2&
E_3&
E_4&
E_5&
E_6&
E_{01}
\cr\tablerule&
6\,x_1 + r\,x_2\ + q\,{x_6^3} & 4\,x_2 + p\,{x_6^2}&u&v&w&0&-12\cr\tablerule&
3\,x_1 + c\,x_2\,x_6 + b\,{x_6^3} & 2\,x_2 + a\,{x_6^2}&0&0&0&x_6&-6\cr\tablerule  
}} \addtocounter{equation}{1}  \eqno(\arabic{equation})$$

\noindent where $p,$ $r,$ $q,$ $a,$ $b$ and $c$ are appropriate
polynomials in the coordinates of the base.

One can find the general solution of the constraint (\ref{EF_constraint}) 
by writing all the polynomials in (\ref{unsound}) and
(\arabic{equation})
as certain expressions in terms of $f,~g,~k_3,~a,~b,~c$. The spectral 
cover is given by the expression in the previous subsection, with $a_0$ and 
$a_2$ evaluated for these constrained polynomials. Making the relevant substitution one finds that 
$a_2$ vanishes identically (thus turning $\Sigma$ into a degenerate cover) 
and that $a_0$ is given by the rather formidable expression :
\bearray
a_0 &=&  243\,{a^6} - 432\,{a^3}\,{b^2} + 192\,{b^4} +
        432\,{a^4}\,b\,c - 384\,a\,{b^3}\,c -
        108\,{a^5}\,{c^2} + 288\,{a^2}\,{b^2}\,{c^2} - \nonumber \\
&\,&         96\,{a^3}\,b\,{c^3} + 12\,{a^4}\,{c^4} +
1944\,{a^4}\,f - 1728\,a\,{b^2}\,f +
        1728\,{a^2}\,b\,c\,f - 432\,{a^3}\,{c^2}\,f +
   	\nonumber \\     
&\,&3888\,{a^2}\,{f^2} - 
 3888\,{a^3}\,g +
        3456\,{b^2}\,g - 3456\,a\,b\,c\,g +
        864\,{a^2}\,{c^2}\,g - 15552\,a\,f\,g + \nonumber \\
&\,&        15552\,{g^2} - 
486\,{a^4}\,k_3 + 
    432\,a\,{b^2}\,k_3 -
        432\,{a^2}\,b\,c\,k_3 +
        144\,{a^3}\,{c^2}\,k_3 -
        32\,{b^2}\,{c^2}\,k_3 + \nonumber \\
&\,&        32\,a\,b\,{c^3}\,k_3 - 
        8\,{a^2}\,{c^4}\,k_3 -
   1944\,{a^2}\,f\,k_3 +
        144\,a\,{c^2}\,f\,k_3 + 3888\,a\,g\,k_3 -
        288\,{c^2}\,g\,k_3 + \nonumber \\
&\,&        324\,{a^2}\,{k_3^2} + 
        96\,b\,c\,{k_3^2} - 
         48\,a\,{c^2}\,{k_3^2} +
        432\,f\,{k_3^2} - 72\,{k_3^3} 
\eearray
This shows rather explicitly how imposition of a sufficient number of fermionic
symmetries in the underlying linear sigma model forces the cover to become 
degenerate. 
The spectral cover is still given by 3 copies of the zero set of the elliptic 
fibration and a posible set of vertical components.

\subsection{A model over a Calabi-Yau threefold realized as a 
hypersurface in a resolution of $\W\P^{12,8,2,1,1}$}

All of the previous models were deformations of the tangent 
bundle. Since the tangent bundle will generally lead to a 
degenerate spectral cover, one should attempt to construct 
models on bundles which are not deformations of $TZ$. The 
model we consider here is of this type, being perturbatively 
well-defined {\em without} any fermionic symmetries. This is 
a radical departure from the tangent bundle, but, as we will see, 
the resulting spectral cover is still degenerate.

The model is constructed over a degree 
 $(24,12,6)$ hypersurface in 
a resolution of $\W\P^{12,8,2,1,1}$. 
The relevant fields are 8 bosonic ($x_1, .., x_7, p$) and
7 fermionic ($\lambda_1, .., \lambda_6, \gamma$) under 3 $U(1)$ 
charges, given by:

\medskip \ifx\answ\bigans\else\hskip-.25in\fi
\hbox{
\qquad
\vbox{\offinterlineskip \tabskip=0pt
\halign{
#&
\vrule height 10pt depth 5pt
\enskip\hfil$#$\hfil\enskip\vrule &
\enskip\hfil$#$\hfil\enskip\vrule &
\enskip\hfil$#$\hfil\enskip\vrule &
\enskip\hfil$#$\hfil\enskip\vrule &
\enskip\hfil$#$\hfil\enskip\vrule &
\enskip\hfil$#$\hfil\enskip\vrule &
\enskip\hfil$#$\hfil\enskip\vrule &
\enskip\hfil$#$\hfil\enskip\vrule &
\enskip\hfil$#$\hfil\enskip\vrule \cr\tablerule&
x_1&
x_2&
x_3&
x_4&
x_5&
x_6&
x_7&
p\cr\tablerule&
12&8&2&1&1&0&0&-24\cr\tablerule&
6&4&1&0&0&1&0&-12\cr\tablerule&
3&2&0&0&0&0&1&-6\cr\tablerule
}}
\qquad \qquad
\vbox{\offinterlineskip \tabskip=0pt
\halign{
#&
\vrule height 10pt depth 5pt
\enskip\hfil$#$\hfil\enskip\vrule &
\enskip\hfil$#$\hfil\enskip\vrule &
\enskip\hfil$#$\hfil\enskip\vrule &
\enskip\hfil$#$\hfil\enskip\vrule &
\enskip\hfil$#$\hfil\enskip\vrule &
\enskip\hfil$#$\hfil\enskip\vrule &
\enskip\hfil$#$\hfil\enskip\vrule \cr\tablerule&
\lambda_1&
\lambda_2&
\lambda_3&
\lambda_4&
\lambda_5&
\lambda_6&
\gamma\cr\tablerule&
1&1& 2&4&6&8&-22\cr\tablerule&
1&1&0&2&3&4&-11\cr\tablerule&
0&0&0&1&2&3&-6\cr\tablerule
}}}

A suitable choice of polynomials which allow for a transverse 
bundle over a smooth threefold is given by:

\bearraynn
G & = & \lambda x_1^2 + \mu x_2^3 + \nu x_3^{12} x_7^6 + (\rho_1 x_4^{24} + 
\rho_2 x_5^{24}) x_6^{12} x_7^6 \\
F_1 & = & (a_1 x_4^{21} x_6^{10} + b_1 x_5^{21} x_6^{10}) x_7^6 \\
F_2 & = & (a_2 x_4^{21} x_6^{10} + b_2 x_5^{21} x_6^{10}) x_7^6 \\
F_3 & = & a_3 x_3^{10} x_6 x_7^6 \\
F_4 & = & a_4 x_3^3 x_1 x_7^2 \\
F_5 & = & a_5 x_2^2 \\
F_6 & = & a_6 x_3^7 x_7^3
\eearraynn

The base is a Hirzebruch surface $F_2$ with homogeneous 
coordinates $x_3,x_4,x_5,x_6$ and the fibre is a sextic in 
$WP_{3,2,1}$.
There are no $E$'s required in this model, and the bundle has 
rank 5. Computation by the methods of section 4 gives  
the spectral cover $ a_4 a_6^3 x_3^{24} x_7^5=0$.

This is again fully degenerate, consisting of 5 
copies of the section of the elliptic fibration of $Z$ plus 
extra vertical components. One can also consider various 
modifications of this model, by picking other solutions 
of  the transversality constraints. This leads to various 
families of $(0,2)$ linear sigma models. For some of these 
families the restriction of $V$ to the generic elliptic fibre of 
$Z$ fails to be semistable, while on others the restriction is 
semistable but leads to a degenerate spectral cover. While we 
have explored a few of these families, we lacked the computational 
power to undertake a complete study.

\section{Discussion}

What are we to make of these results? Both the physical and the
mathematical answers appear to depend in part on the dimension
of the compactification under study. From a mathematical
perspective, the moduli space of (semi-) stable vector 
bundles over a $K3$ surface differs
significantly from that which arises over a Calabi-Yau $3$-fold.
In the case of the latter, it is believed that the moduli space 
(for a fixed Calabi-Yau base)
is {\it stratified} into various components. In some of these
components, it is expected that the spectral cover is
of the degenerate form that we have found. On such loci,
the moduli of the bundle over the Calabi-Yau are associated
with moduli of the spectral bundle over the spectral cover.
This is the situation in which we have found ourselves
regarding our $(0,2)$ linear sigma model constructions.
On the other hand, the mathematical situation is quite 
different over a $K3$ surface. The moduli space has no analagous
stratification and therefore the generic deformation
away from a bundle with a degenerate spectral cover takes
us to a bundle with a non-degenerate cover.

As far as our calculations go, this means that if we were to include
a generic deformation in $K3$ examples (assuming that such a generic
deformation can be torically represented) the spectral
cover should split apart. But in our Calabi-Yau $3$-fold examples,
we would expect --- as we have found --- that the cover
may remain degenerate.

A natural question, though, is why --- in both the $K3$ and
Calabi-Yau $3$-fold cases --- have we ``landed'' at a point
with a degenerate spectral cover. We suspect the reason to lie
in the fact that the linear sigma model is a physical
representation of toric geometry. Toric constructions ---
although quite useful --- are also quite special. The central
element of linear $\C^*$ actions might very well account for
the degeneracy we find.

From a physical point of view, it is also the case that
compactifications on $K3$ surfaces and on Calabi-Yau $3$-folds
have quite different properties. This is most easily
gleaned in the dual $F$-theory picture in which we
are comparing $F$-theory with six and with four non-compact
dimensions, respectively. In  four dimensions --- and not
in six --- it was shown by \cite{SVW} that the $F$-theory vacuum 
will be unstable, if the Calabi-Yau $4$-fold on
which the compactification is done 
has nonzero Euler number. This is due to the appearance of a nonzero
one-point function for the four-form field. 
This means that the model is inconsistent unless one includes 
a certain number, $I_i$, of three-branes to cancel up the tadpole. 
The worldvolume of such three-branes is taken to fill out the 
uncompactified spatial directions, in order to preserve the 4-dimensional 
Poincare invariance of the vacuum.
As shown in \cite{SVW}, in the case of a {\em smooth}
Weierstrass model $X$, one must have:
\be
I_i=\chi(X)/24.
\ee
Under duality, these F-theory three-branes are conjectured
to map to $I_i$ heterotic five-branes (called 5-brane `defects' in what 
follows) which are wrapped over certain elliptic 
curves of $Z$. 

Therefore, it would appear that the dual heterotic models for
F-theory on a fourfold $X$ with nonzero Euler number necessarily include
such 5-branes. In fact, this reasoning is a little too quick, as discussed 
for the first time in \cite{BJPS}. 
To understand why, one can start with the case of $6$-dimensional 
compactifications (which is better understood) and use an 
`adiabatic argument' to descend to $4$ dimensions.

Recall that heterotic compactifications on a K3 surface 
admit degenerate limits in which `small instantons' are formed. Such a limit
can be understood as a degeneration of the 
heterotic bundle $V$ to a torsion-free sheaf ${\cal E}$. In general, 
such a sheaf has a singularity locus of codimension at most two, so in the K3 
case the sheaf fails to be locally free at a finite number of points. 
On the other hand, via the Uhlenbeck-Yau theorem, one can think about this 
in terms of $SU(r)$ anti-self-dual Yang-Mills 
connections, i.e. instantons on the 
underlying real 4-manifold. In this case, the degeneration above can be 
interpreted as the limit when certain instantons collapse to zero size. 
Such small instantons can be also be 
interpreted as 5-branes whose worldvolumes 
fill the 6 uncompactified directions and intersect the K3 surface at the 
singular points of ${\cal E}$. 

It is natural to assume that a similar picture holds in 4-dimensional 
compactifications (over a Calabi-Yau 3-fold $Z$). More precisely, 
there will exist degenerations ${\cal E}$ 
of $V$ (presumably to some type of torsion-free sheaves), whose singularity 
locus will be a subvariety of codimension 2 in $Z$, i.e. an algebraic curve
$\Gamma$ embedded in $Z$. There exist at least two interesting cases to 
consider, namely when $\Gamma$ is the lift of a curve $\gamma$ in the base 
$B_H$ (i.e. the image of a section of $X|_\gamma$) and when $\Gamma$ coincides 
with an elliptic fibre of $Z$. Such singularizations of $V$ can again be 
interpreted as heterotic compactifications containing 
$5$-branes wrapped over $\Gamma$.

To motivate these ideas, one can imagine `fibering' the
6-dimensional picture above over a base of complex dimension 1. This will 
give a degeneration of $V$ over a curve of the first type, whose 
associated 5-branes we will call `instantonic' 5-branes. On the other 
hand, it is natural to identify the second type of 5-branes with the 5-brane
defects discussed above. Both these types of 5-branes will bring a nontrivial 
contribution to the anomaly cancellation conditions on the heterotic side, 
which becomes:
\be
c_2(V)+I_d +I_i =c_2(TZ)
\ee

The inverse process --- desingularizing ${\cal E}$ to a stable bundle $V$ --- 
can be interpreted as a 5-brane `dissolving' into a gauge instanton of finite 
size. If the above picture indeed holds for both types of 5-branes, such a 
process should be possible not only for instantonic 5-branes, but also for 
5-brane defects. On the $F$-theory side, it should therefore be possible to 
pass from a situation in which 3-branes are present to a situation with no 
3-branes. How is such a process realized ? As argued in \cite{BJPS}, this 
should correspond to the F-theory 3-branes `dissolving' into some  
coincident 7-branes, i.e. expanding to become finite size instantons 
of the worldvolume theory of those 7-branes. It was argued in \cite{BJPS} 
that a 3-brane can dissolve only on a multiple 7-brane,
i.e. a state of at least two 7-branes sitting on top of each other. Such 
multiple 7-branes are partially wrapped over components of the discriminant 
locus of $X$ at which the discriminant vanishes in order at least 2; 
therefore, they will correspond to loci above which $X$ becomes singular. 
As the 3-branes expand to finite size instantons, part of the associated 
enhanced gauge group will be broken. In such a situation, therefore, the 
enhanced gauge symmetry on the F-theory side will be less than that predicted 
by Tate's algorithm. Similarly, the gauge symmetry will be reduced on the 
heterotic side since the dual transition of a 5-brane to a finite size 
instanton increases the holonomy of the heterotic bundle.

We see therefore that one can have F-theory compactifications with
$\chi(X)\neq 0$ but with no 3-brane defects. In this situation the tadpole 
will be canceled by the presence of instantons on the 7-brane and the 
F-theory consitency condition is modified to :
\be
c_2(S)=\chi(X)/24
\ee
where $S$ is the gauge bundle over the 7-branes.

Since the spectral cover of $V$ is controlled by the complex structure of $X$,
it follows that the moduli of $S$ must map to the moduli of the spectral 
bundle $W$ of $V$. $W$ is a bundle over the spectral cover of $V$, which 
generalizes the line bundle $L\rightarrow \Sigma$ to the case when $\Sigma$ 
has multiple components.

Now, if $S$ is a nonsplit higher
rank bundle, then the same shoud be true of $W$. 
But the only way that this can happen is if
the spectral cover is degenerate so that the line bundles
$L_i$ from merging components can themselves merge into
a higher rank bundle. Therefore, the heterotic dual of an F-theory model 
with $\chi(X) \neq 0$, with no 3-branes and with an irreducible instanton 
connection (more precisely, not fully reducible) on the multiple 7-branes 
is expected to have a 
degenerate spectral cover. This is precisely the situation we have found 
in all of the models we studied. 

	To understand why this happens for this class of models,
remember that we carefully chose our $(0,2)$ model data such that the 
heterotic compactifications be consistent as purely perturbative theories 
(and in particular satisfy 
$c_2(V)=c_2(TZ)$); it follows that in our case there are no 5-branes on the 
heterotic side. Therefore, F-theory on $X$ cannot contain 3-branes.
In all of our models $Z$ was also 
chosen to have an `obvious' elliptic fibration. 
In the spirit of \cite{FMW}, we therefore expect that their F-theory duals 
are realizable on fourfolds $X$ presented as Weierstrass models in some toric 
varieties.
Typically such an $X$ admits a desingularization 
(obtained by varying its complex structure) to a smooth 
Weierstrass model $X_{smooth}$ realized in the same way.  
Then a computation similar 
to that of of \cite{SVW} will generally imply $\chi(X_{smooth})>0$. 
Thus the F-theory 
vacuum on $X_{smooth}$ will be forced to contain 3-brane defects. 
As we singularize 
to obtain $X$, the number of these 3-branes can decrease only if some of 
them dissolve into some multiple 7-branes. Therefore, the only way that 
$F$-theory on $X$ will have no 3-brane defects is that all of these 3-branes 
have dissolved into the 7-branes. 

These rather abstract arguments lead us to believe that the reason for 
obtaining completely degenerate spectral covers is that the condition that 
$Z$ have an obvious fibration essentially forces $X_{smooth}$ to have a 
nonzero Euler number and that the vacuum configuration of the gauge theory 
on the corresponding multiple 
$F$-theory 7-branes is given by a gauge connection which is not fully 
reducible.
We now proceed to partially test this hypothesis in our 4-dimensional examples.
Since we do not have enough control of the gauge connections on the 7-branes, 
the best we will be able to do here is to test whether the Euler number of 
$X_{smooth}$ is nonzero.

\subsection{Determination of the dual fourfols}

\subsubsection{The $F$-theory duals of the models of section 5.2}

Let us first concentrate on the examples of subsection  
5.2. The rank of the bundle is 3 and the ten dimensional $E_8 \times E_8$ 
gauge  group is broken to $E_8 \times E_6$.

Having an explicit expression for the spectral cover one can   
attempt to construct the $F$-theory dual along the lines of \cite{FMW}.
Although the work of \cite{FMW} was concerned only with the case of generic
spectral covers, let us ignore possible subtleties
\footnote{Subtleties occur due to the following reason: since we are 
considering a degenerate cover and since the spectral bundle need not
be a direct sum of line bundles over the components, the argument employed in
\cite{FMW} to determine the dual fourfold may have to be modified.}
associated to the 
degenerate character of the cover and see where direct 
application of \cite{FMW} leads us.

In the examples of subsection 5.2 the heterotic base is a $\P^2$, 
with anticanonical line bundle 
$K^{-1}_{B_H}=O_{\P^2}(3)$. The toric data is the same for all of the models 
in that subsection. The heterotic 3-fold is a Weierstrass model embedded 
in the ambient space 
$\P_{2,3,1}(O_{B_H}(6)\oplus O_{B_H}(9)\oplus {\cal O}_{B_H})$. In the 
notation of section 2, the fibre variables are $x:=x_2,~y:=x_1,~z:=x_6$,
which are sections of $K_{B_H}^{-2}\otimes O_{\P_H}(2)\approx O_{\P_H}(6,2), 
K_{B_H}^{-3}\otimes O_{\P_H}(3)\approx O_{\P_H}(9,3)$ and 
$O_{\P_H}(1)\approx O_{\P_H}(0,1)$ respectively. 
Homogeneity constrains the general spectral cover for this toric data to be of 
the form :
\begin{equation}
{}~ a_0z^3 + a_2xz + a_3y=0~,
\end{equation}
where $a_0, a_2$ and $a_3$ are polynomials over $B_H$ of degrees
$d, d-6$ and $d-9$. Since our models are purely perturbative we have 
$c_2(V)=c_2(TZ)$.
The precise value of $d$ can be fixed by computing 
$\pi_{H,*}(c_2(TZ))$ along the lines of \cite{SVW}. 
Following the procedure described there one obtains:
\be
\pi_{H,*}(c_2(V))=12 c_1(TB_H)=12 c_1(K^{-1}_{B_H})
\ee 
so that:
\be
\label{N_class}
{\cal N}=K_{B_H}^{-12}
\ee
This relation is valid for a Weierstrass model over an arbitrary toric base
$B_H$. In our case $K_{B_H}^{-1}=K_{\P^2}^{-1}=O_{\P^2}(3)$ and 
we obtain:
\be
{\cal N}=O_{\P^2}(36)
\ee
which fixes $d=36$. 
Note that, since $B_H\approx \P^2$ is not a Hirzebruch surface, we cannot 
apply (\ref{class3}) directly to our model.

Let us first describe the toric data of the $F$-theory dual.  
According to section 2, $X$ is a Weierstrass model in the ambient space 
$\P_F:=P_{2,3,1}(K_{B_F}^{-2}\oplus K_{B_F}^{-3}\oplus {\cal O}_{B_F})$
where $B_F$ is the ruled 3-fold 
$\P({\cal M}\oplus O_{B_H})$ with ${\cal M}$ a line bundle over 
$B_H$. Since $B_H\approx \P^2$, we have ${\cal M}=O_{\P^2}(n)$ for some 
$n \in \Z$. Then $K_{B_F}^{-1}=K_{B_H}^{-1}\otimes O_F(2)\otimes {\cal M}=
O_{B_H}(n+3,2)$. 
The fibre coordinates $X,Y,Z$ are sections of the line bundles 
$K_{B_F}^{-2}\otimes O_{\P_F}(2)\approx O_{P_F}(2n+6,4,2), 
K_{B_F}^{-3}\otimes O_{\P_F}(3)\approx O_{P_F}(3n+9,9,3)$ and 
$O_{\P_F}(1)\approx O_{P_F}(0,0,1)$. Denoting the homogeneous coordinates 
of the projectivization $B_F=\P({\cal M}\oplus O_{B_H})$ (essentially the 
homogeneous coordinates of the $\P^1$ fibre of $B_F \rightarrow B_H$) by 
$u,v$ as in section 2, the toric data for the dual fourfold $X$ is :
$$
\medskip\centerline{
\vbox{\offinterlineskip \tabskip=0pt
\halign{
#&
\vrule height 10pt depth 5pt
\enskip\hfil$#$\hfil\enskip\vrule &
\enskip\hfil$#$\hfil\enskip\vrule &
\enskip\hfil$#$\hfil\enskip\vrule &
\enskip\hfil$#$\hfil\enskip\vrule &
\enskip\hfil$#$\hfil\enskip\vrule &
\enskip\hfil$#$\hfil\enskip\vrule &
\enskip\hfil$#$\hfil\enskip\vrule &
\enskip\hfil$#$\hfil\enskip\vrule \cr\tablerule&
x_3&
x_4&
x_5&
u&
v&
X&
Y&
Z \cr\tablerule&
1&1&1&n&0&2n+6&3n+9&0\cr\tablerule&
0&0&0&1  &1&4&6&0\cr\tablerule&
0&0&0&0  &0&2&3&1\cr \tablerule
}
}
}
$$
$X$ is given by the zero locus of the section $Y^2-X^3-FXZ^4-GZ^6 \in 
H^0(K_{B_F}^{-6}\otimes O_{\P_F}(6))=H^0(O_{\P_F}(6n+18,12,6))$ where 
$F,G$ are sections of $K_{B_F}^{-4}=O_{B_F}(4n+12,8)$ and 
$K_{B_F}^{-6}=O_{B_F}(6n+18,12)$. The expansions 
(\ref{F_expansion},\ref{G_expansion}) give coefficients $F_i,~G_j$ which can 
be viewed as polynomials over $B_H$ of degrees $(4-i)n+12$, respectively 
$(6-j)n+18$.

According to \cite{BIKMSV}, to identify $F_i,~G_j$ we 
must first determine the unbroken gauge symmetry of the model, which in our 
case is $E_6 \times E_8$. Then $X$ should have a section of $E_6$ 
singularities (which we can place at $u=0$) and a section of $E_8$ 
singularities (placed at $v=0$). Via Tate's algorithm 
\cite{BIKMSV}, this forces all $F_i,~G_j$ to be zero except for 
$G_4,~G_5,~G_6,~G_7$ 
and $F_3,~F_4$. It also requires that $G_4={\tilde a}_3^2$ for some polynomial
$a_3$ of degree $n+9$ over $B_H$. Therefore, the defining equation of $X$ 
takes the form :
\be
Y^2=X^3+({\tilde a}_2 u^3 v^5+ F_4 u^4 v^4)XZ^4 + 
({\tilde a}_1^2 u^4 v^8 + {\tilde a}_0 u^5 v^7 + G_6 u^6 v^6+G_7 u^7 v^5)Z^6
\ee
where we denoted $F_3,~G_5$ by ${\tilde a}_2, {\tilde a}_0$.
In this equation, $F_4$ and $G_6$ are controled by the complex structure of 
$Z$ while 
$F_3 ={\tilde a}_2 \in H^0(O_{B_H}(n+12)),
~{\tilde a}_3\in H^0(O_{B_H}(n+9))$ and 
$G_5={\tilde a}_0\in H^0(O_{B_H}(n+18))$ are controled by the heterotic 
spectral cover $\Sigma$. There is a simple way to determine $n$ directly 
if one notes 
that, on the heterotic side, there is no extra matter transforming 
in a representation of the first $E_8$. Then the results of \cite{BIKMSV,KV}
imply that $G_7$ cannot have any zeroes, which fixes $n=18$. Alternatively, 
this follows from the conjecture of \cite{FMW}, according to which $a_j$ 
and ${\tilde a}_j$ should be identified. This fixes the value of $n$ to $18$ 
and partially determines the equation of $X$. 

Now let us return to the perturbatively consistent examples of subsection 5.2.
In those cases the spectral cover is degenerate ($a_2=a_3=0$), 
so that ${\tilde a}_2={\tilde a}_3=0$. This gives $X$ in the form: 
\be
Y^2=X^3+F_4 u^4 v^4XZ^4 + 
({\tilde a}_0 u^5 v^7 + G_6 u^6 v^6 +G_7 u^7 v^5)Z^6
\ee
According to Tate's algorithm, this corresponds to a section of  
$E_8 \times E_8$ singularities of $X$. Naive application of the work 
of \cite{BIKMSV} would then lead to the conclusion that $F$-theory on $X$
has an unbroken $E_8 \times E_8$ gauge symmetry. This contradicts the 
$E_8 \times E_6$ gauge symmetry that we find perturbatively 
on the heterotic side ! How can we understand this apparent discrepancy ?
The only explanation we envisage is the possibility, discussed above, 
of having an F-theory 
compactification whose vacuum contains nontrivial 
instanton configurations of the effective gauge theory of its multiple 
7-branes. Such configurations  break the
effective gauge group from $E_8\times E_8$ to $E_8 \times E_6$. 

To lend more credence to our  hypothesis, let us compute the Euler number 
of the {\em generic} member $X_{gen}$ 
of the family of Weierstrass models to which $X$ 
belongs. Such a member is smooth and given by a Weierstrass equation  
of the form $Y^2-X^3-FXZ^4-GZ^6~=0$ in $\P_F$, the 
polynomials $F,~G$ being generic with the same multidegrees as above.
The computation of $\chi(X_{gen})=c_4(X_{gen})[X_{gen}]$ can be achieved by 
the methods of \cite{SVW} and gives :
\be
\label{chi}
\chi(X_{gen})=12c_1(c_2 + 30 c_1^2)[B]
\ee
where $c_j:=c_j(TB_F)$.
In our case $TB_F$ is given by the exact sequence :
\be
\label{adjunction}
0 \longrightarrow O_{B_F}^{\oplus 2} \longrightarrow 
O_{B_F}(1,0)^{\oplus 3}\oplus O_{B_F}(0,1)\oplus O_{B_F}(1,n)\longrightarrow 0
\ee
which allows us to express $\chi(X_{gen})$ in terms of triple intersections  
of the toric divisors of $B_F$. $H^2(B_F,\Z)$ is generated by the 
classes
$\Delta_x=[D_{x_3}]=[D_{x_4}]=[D_{x_5}],~\Delta_u:=[D_u],~\Delta_v:=[D_v]$
of the toric divisors $D_{x_i}=(x_i),~D_u:=(u),~D_v:=(v)$. We have 
$c_1(O_{B_F}(1,0))=\Delta_x,~c_1(O_{B_F}(n,1))=\Delta_u,
~c_1(O_{B_F}(0,1))=\Delta_v$. This gives the relation 
$\Delta_u=n\Delta_x+\Delta_v$. 
The triple intersections can be expressed torically 
as mixed volumes of polytopes 
and are given by $\Delta_x^3=0,~\Delta_x^2\Delta_v=1,~\Delta_x\Delta_v^2=-n,
~\Delta_v^3=n^2$. Using this, an easy computation gives the number of 3-branes:
\be
\chi(X_{gen})/24=30n^2+822
\ee
which is positive, as expected. For $n=18$, it is equal to $10542$.

\subsubsection{The $F$-theory dual of a purely perturbative model with 
$B_H=F_n$}

The model of subsection 5.3 has $B_H=F_2$. It turns out that the generic 
member of the family of its $F$-theory 
dual also has a positive Euler number. In fact, we now show that this will be 
the case for any purely perturbative heterotic model with $B_H$ a generalized
Hirzebruch surface $F_n$. For such a model we have ${\cal N}=K^{-12}_{F_n}=
O_{F_n}(12(n+2),24)$. The dual is specified by the line bundle ${\cal M}=
O_{F_n}(m,k)$. Then $B_F=\P({\cal M}\oplus {\cal O}_{F_n})=F_{n,m,k}$
(a generalized Hirzebruch). According to the conjecture of \cite{FMW}, 
the line bundle ${\cal M}$ will be of the form ${\cal M}={\cal N} \otimes 
K^{6}_{F_n}$. Then 
${\cal M}=K^{-6}_{F_n}=O_{F_n}(6(n+2),12))$. 
Therefore $m=6(n+2)$ and $k=12$. The Euler 
number of a smooth Calabi-Yau Weierstrass model with base $F_{n,m,k}$ was 
computed in \cite{BJPS} and is given by:
\be
\chi(X_{gen})/24=732+60km-30k^2n
\ee
In our case, this gives:
\be
\chi(X_{gen})/24=9372
\ee
which is positive and independent of $n$.

\section{Conclusions}

We presented a method for computing the spectral cover 
associated to heterotic compactifications which are realizable via 
$(0,2)$ linear sigma models over an elliptically fibered Calabi-Yau manifold. 

Contrary to naive expectations, we discovered that the most accessible 
models give rise to degenerate covers of a rather 
trivial form. This indicates that in such cases most of the information 
of the bundle is translated to instanton moduli on multiple $F$-theory 
7-branes, as suggested in \cite{BJPS}. We thus found indirect evidence 
that a large number of purely perturbative heterotic compactifications 
are dual to $F$-theory vacua realized on singular limits of smooth 
Calabi-Yau fourfolds with nonzero Euler number,  
containing (in the limit) instanton configurations on 
multiple 7-branes but no 3-brane defects.

The $F$-theory encoding 
of the moduli space of such heterotic models is therefore not completely 
described by the geometry of the dual 4-fold; rather, a large piece of 
information is again specified by 
gauge connections. This opens the door to new `sub-duality' 
conjectures, which seem to be intimately related to topological field 
theories and integrable systems. From a mathematical point of view, 
such models draw attention to the relevance of degenerate spectral covers 
and the associated bundles, an issue which as yet has not been intensively 
studied.

Due to computational difficulties, 
we avoided considering more general models 
(with a `non-obvious' fibration structure). It is an interesting problem to 
generalize the methods presented here to such situations. It should be noted 
that elliptically fibered Calabi-Yau fourfolds of zero Euler number 
are in a certain sense rather rare \cite{SVW,KLRY}.

As for the story in $6$ dimensions, 
it is important to understand better 
how the spectral bundle moduli are realized. 
A detailed
analysis seems to require understanding the moduli 
space of a certain class of sheaves over the nonreduced scheme associated to 
such a cover. Fully describing the corresponding $F$-theory duals 
would require an analysis of certain 
`twist' moduli of the effective field theory of partially wrapped 
multiple 7-branes. 
We hope to report on these and related problems in a future publication.

\bigbreak\bigskip\bigskip\centerline{{\bf Aknowledgements}}\nobreak
\bigskip

We are grateful to  Ron Donagi, Robert Friedman and John Morgan for 
insightful discussions. 
M.~B. and A.~J. wish to thank Peter Mayr, Tony Pantev and Vladimir Sadov
for collaboration during the early stage of this work.
The research of M.~B.~ and A.~J.~ was partially supported by the
NSF grant PHY-92-18167, the NSF 1994 NYI award and the  DOE 1994 OJI
award. T.~M.~C. is supported by the National Science Foundation. 
B.~R.~G. is supported by a National Young Investigator Award and an 
Alfred P. Sloan Foundation Fellowship. C.~I.~L. is supported by a C.U. 
Fister Fellowship. 
The work of B.~R.~G. and C.~I~.L. is also supported by the DOE grant 
DE-FG02-92ER40699B.

\vspace{0.2in}

\appendix
\section{Some classical results on holomorphic vector 
bundles over nonsingular elliptic curves}

Let $E$ be a nonsingular elliptic curve, together with a distinguished point 
$p \in E$.
Topological vector bundles $V$ over $E$ are classified 
 by the pair $({\rm rank}V,{\rm deg}V)$, where 
deg$V:={\rm deg}({\rm det}V)$.
Here ${\rm det}V:=\Lambda^{{\rm rank}V}(V)$ is the
determinant bundle of $V$. In the holomorphic category (which is
more `rigid') the classification is finer. 

Consider first the case of line bundles. Let $\Pic^d(E)$ be the set 
of line bundles of degree $d$ over $E$.
If $L$ is a degree 
zero holomorphic line bundle over $E$, then there exists a unique 
point $q \in E$ such that $L\approx O_E(q-p)$. This gives a 
$p$-dependent isomorphism $\Pic^0(E) \approx E$.
If $L \in \Pic^d(E)$, then $L \otimes O_E(-dp)\in \Pic^0(E)$ so we can 
write $L\approx O_E(q+(d-1)p)$ for a uniquely determined $q \in E$.
Thus, we have $p$-dependent isomorphisms $\Pic^d(E) \approx E$ 
for all $d \in \Z$.

	The case of higher rank holomorphic vector bundles is more
complicated. First, the
Riemann -Roch theorem for an elliptic curve states that for any
holomorphic vector bundle $V$ over $E$ we have
$\chi(V):=h^0(V)-h^1(V)={\rm deg}V$.
	Further, for any such $V$ we can consider a maximal decomposition
into {\em holomorphic} subbundles:  $V=V_1\oplus ...\oplus V_k$. This
reduces the problem to the classification of holomorphic {\em
indecomposable} vector bundles of fixed rank $r$ and fixed degree
$d$, whose set of isomorphism classes we denote by 
${\cal E}(r,d)$.

	To study this problem it proves useful to first look at the spaces of 
sections $H^0(V)$ of such bundles. Atiyah \cite{Atiyah} 
proves the that for any $V \in {\cal E}(r,d)$ we have
$h^0(V)=d$,  if $d > 0$, respectively $h^0(V) \in \{0,1\}$, 
if $d =0$. Moreover, there exists a unique bundle $F_r \in {\cal E}(r,0)$ 
(up to isomorphism) such that $h^0(F_r) \neq 0$ and we have $h^0(F_r)=1$.
The bundles $F_r$ can be constructed inductively by taking
successive nontrivial extensions of $O_E$ by itself. That is, $F_1:=O_E$
and for any $r > 1$, $F_r$ is the unique nonsplit extension of $F_{r-1}$
by $O_E$. In particular, the bundles $F_r$ have trivial determinant 
and are semistable.

	First consider the case $d=0$, which is our primary focus in the
	present paper. Atiyah shows that any indecomposable 
	holomorphic vector bundle $V \in {\cal E}(r,0)$ can be written
	in the form $V=L\otimes F_r$ with $L \in \Pic^0(E)$ a 
uniquely-determined vector bundle (which obviously satisfies $L^r=\det V$).
This shows that ${\cal E}(r,0)\approx \Pic^0(E)$.
Thus, there is essentially no more `content' in degree zero indecomposable
vector bundles then there is in degree zero line bundles.

The explicit description of ${\cal E}(r,d)$ for $d>0$ is a classical result
of Atiyah \cite{Atiyah}, which states that, given a pair $(E,p)$,
any $V \in {\cal E}(r,d)$ is of the
form $V=L\otimes E(r,d)$,
where $E(r,d)$ is a special element of ${\cal E}(r,d)$ (defined up to
isomorphism) and $L$ is a line bundle of degree zero on $E$.
Here $L$ is defined up to tensoring by an arbitrary holomorphic line
bundle of order $r/h$ on $E$, where $h:={\rm gcd}(r,d)$. 
The bundles $E(r,d)$ are
defined (up to isomorphism) by an inductive procedure
involving extensions by trivial vector bundles and tensoring with
$O(p)$. In particular, one has $E(r,0)=F_r$ for all $r \geq 0$.  The
bundles $E(r,d)$ have ${\rm det}E(r,d) = O(d\times p)$. 
Using this result, one can show that ${\cal E}(r,d)$ is in bijection with 
$E$. More precisely, there is a ($p$-dependent) bijection from ${\cal
E}(r,d)$ to ${\cal E}(h,0)$ and a bijection from ${\cal E}(h,0)$ to
$\Pic^0(E)$, the last map being given by $V \rightarrow L$ where $L \in
\Pic^0(E)$ is uniquely determined (up to isomorphism) by
$V=L\otimes F_h$, as explained above. Via the usual $p$-dependent
isomorphism $ \Pic^0(E) \approx E$, this induces the desired identification.
The description of ${\cal E}(r,d)$ for $d<0$ can be obtained by
dualizing the above.

	If one asks for a suitable moduli space of vector bundles over a
Riemann surface of genus $g$, 
one finds that it is appropriate to restrict the set
of bundles in order to obtain a reasonable result.
The correct notion is that of semistable vector bundles.
For any holomorphic vector bundle $V$, define its 
{\em normalized degree} $\mu(V)$ by $\mu(V):=\deg V/\rank V$. 

Then $V$ is called {\em semistable} if $\mu(W) \le \mu(V)$
for all proper subbundles $W$ of $V$. 
If strict inequality holds for all such $W$, then $V$ is called 
{\em stable}.

To construct the moduli space $M(r,d)$ of semistable bundles of rank $r$ and 
degree $d$ over a Riemann surface of genus $g$, 
one considers the set $ss(r,d)$ of isomorphism classes 
of such bundles and divides it by an equivalence relation called 
$S$-equivalence. To define this relation, one first shows that 
any semistable $V$ admits a `Jordan-Holder' filtration :
\be
\label{filtration}
0=V_k\subset V_{k-1}\subset ... \subset V_0=V
\ee
by semistable subbundles $V_j$ with $\mu(V_j)=\mu(V)$ and 
with the property that each succesive 
quotient $V_j/V_{j+1}$ is stable and of normalized degree 
$\mu(V_j/V_{j+1})=\mu(V)$. Although such a filtration is not unique, the
isomorphism class of the associated graded bundle 
$gr(V) := \oplus_{j=0..k-1}{V_j/V_{j+1}}$
does not depend on its choice. If $V$ is stable, then the only such 
filtration is of the form $0=V_1\subset V_0=V$ and in this case 
$gr(V)\approx V$

The $S$-equivalence relation on $ss(r,d)$ is defined by  
$V_1 \equiv V_2 $ iff $gr(V_1) \approx gr(V_2)$. 
If $V_1$ is stable, then $V_1 \equiv V_2$ iff $V_1\approx V_2$.

The desired moduli space is then very roughly given by :
\be
\nonumber
M(r,d) :=ss(r,d) /\equiv 
\ee

More precisely, $M(r,d)$ is constructed as a coarse moduli space 
by using Mumford's geometric invariant theory. $M(r,d)$ is a projective 
variety, whose (closed) points correspond to $S$-equivalence 
classes of semistable bundles. There exists an open subset $M^s(r,d)$ 
of $M(r,d)$, whose points correspond to isomorphism classes of stable 
bundles. All the points of $M^s(r,d)$ are smooth points of $M(r,d)$ 
and the converse also holds unless $g=r=2$ and $d\equiv 0 ({\rm mod} 2)$.
Moreover, if $r$ and $d$ are coprime then $M(r,d)=M^s(r,d)$.
If $M^s(r,d)$ is nonvoid then $\dim M(r,d)=r^2(g-1)+1$.

Now consider the case of an elliptic curve.  
Assume $d\ge 0$ and let $h:={\rm gcd}(r,d)$. 
Then the main results of relevance for us can be summarized as follows :

Each $S$-equivalence class in $ss(r,d)$ contains a unique 
(up to isomorphism) bundle of the form 
$V=\oplus_{i=1}^h{E_i}$ with $E_i$ all stable, of equal rank $r/h$ and
equal degree $d/h$, which is the only member of the 
class for which $V\approx gr(V)$.

In particular:

(01)If $h>1$ 
then there exist no stable bundles of rank $r$ and degree $d$ over $E$.

(02)If $h=1$ then every $S$-equivalence class of semistable bundles over $E$
contains a stable bundle, which is unique in that class up to isomorphism.

(1) $M(r,d)$ is a smooth projective variety of dimension $h$. Moreover, 
there exists an isomorphism $M(r,d) \approx {\rm Sym}^h(\Pic^0(E))\approx 
{\rm Sym}^h(E)$, where 
${\rm Sym}^h$ denotes $h$-th symmetric power. In particular, we 
have $M(r,d) \approx M(sr,sd)$, for any positive integer $s$.

(2) If $h=1$, then the map $\det:M(r,d) \rightarrow \Pic^d(E)$ is an 
isomorphism.

	For semistable vector bundles of degree zero over $E$, the filtration
(\ref{filtration}) takes the form :
\be
0=V_r\subset V_{r-1} \subset ... \subset V_0=V
\ee
where $r:={\rm rank}V$. 
In this case, the subbundles $V_j$ have rank$V_j=r-j$ for all 
$j$ and the associated graded pieces $L_j:=V_{j}/V_{j+1} ( j = 0 ... r-1 ) $
are degree zero line bundles on $E$. 
This gives $gr(V)=\oplus_{j=0..r-1}{L_j}$,  
with $L_j \in \Pic^0(E)$ and the above-mentioned 
bijection $ss(r,0) \approx Sym^r(\Pic^0(E))\approx Sym^r(E)$. 

	Now let $V$ be a rank $r$ and degree zero semistable vector bundle 
over $E$ and let $V':=V\otimes O_E(p)$ for some $p \in E$.  
Then $V$ must have a maximal splitting (direct sum decomposition) of 
the form : 
\begin{equation}
\label{general_splitting}
	V=\sum_{j=1 .. k}{O(q_j -p)\otimes F_{r_j}}
\end{equation}
where $\sum_{j=1..k}{r_j}=r$ 
\footnote{This follows easily by using deg$V=0 $ and the fact 
that $V$ is semistable to show that all terms of a maximal splittting of $V$ 
have degree zero. As such terms are necessarily indecomposable,  
a result quoted above shows that they must be of the form $L_j\otimes 
F_{r_j}$, with $L_j=O(q_j - p)$ some line bundles of degree zero.}. 
It is easy to see 
that the converse is also true, since the direct sum of any finite set of
semistable vector bundles of equal normalized degree is semistable 
(\cite{Seshadri},p17,Cor 7). Thus, we have the following simple fact :

\begin{Proposition}
\label{splitting_proposition}
Let $V$ be a degree zero holomorphic vector bundle over $E$.
Then the following are equivalent :

(a) $V$ is semistable

(b) $V$ has a maximal splitting of the form (\ref{general_splitting})

(c) The twisted bundle $V':=V \otimes O(p)$ has a maximal splitting of the 
form:
\be
\nonumber
V'=\sum_{j=1 ... k}{O(q_j)\otimes F_{r_j}}
\ee
In this case,the {\em multiset} $(r_1,q_1)...(r_k,q_k)$ will be called 
{\em the splitting type } of $V$.

\end{Proposition}

As $O(q_j)\otimes F_{r_j}$ is  indecomposable and of degree $r_j$, 
one of the 
above results shows that $h^0(O(q_j)\otimes F_{r_j}) = r_j$, 
so that, for $V$ {\em semistable} and of degree zero, 
$H^0(V')$ is an $r$-dimensional $\C$-vector space. 
Note that this is true indifferent of the precise splitting type of 
$V$.

\end{document}